\documentclass[a4paper,fleqn,usenatbib]{mnras}

\usepackage{newtxtext,newtxmath}

\usepackage[T1]{fontenc}
\usepackage{ae,aecompl}

\usepackage{graphicx}	
\usepackage{amsmath}	
\usepackage{amssymb}	

\newcommand\hi{H{\small I}}
\newcommand\persqcm{~cm$^{-2}$}
\newcommand\htoo{H$_2$}

\newcommand\atlas{ATLAS$^{\rm 3D}$}
\newcommand\kms{km s$^{-1}$}
\newcommand\solmass{M$_\odot$}
\newcommand\arcdeg{\degr}
\newcommand\mstar{M$_\star$}

\newcommand\mhtoo{M(H$_2$)}
\newcommand\mjb{mJy~beam$^{-1}$}
\newcommand{\e}[1]{$\times 10^{#1}$}

\newcommand{\jykms}{Jy~km~s$^{-1}$}

\defcitealias{serra_vcirc}{S16}
\defcitealias{serra12}{S12}
\defcitealias{serra14}{S14}

\title[HI in Counterrotating Stellar Discs]{Atomic Hydrogen Clues to the Formation of Counterrotating Stellar Discs}

\author[L.~M.~Young et al.]{
Lisa M.\ Young,$^{1,2}$\thanks{E-mail: lisa.young@nmt.edu}
Davor Krajnovi\'c,$^{3}$
Pierre-Alain Duc,$^{4}$
and Paolo Serra$^{5}$
\\
$^{1}$Physics Department, New Mexico Tech, 801 Leroy Place, Socorro, NM 87801 USA\\
$^{2}$Adjunct Astronomer, National Radio Astronomy Observatory\\
$^{3}$Leibniz-Institut f\"ur Astrophysik Potsdam (AIP), An der Sternwarte 16, D-14482 Potsdam, Germany\\
$^{4}$Observatoire Astronomique de Strasbourg, Universit\'e de Strasbourg, CNRS, UMR 7550, 11 rue de
l'Universit\'e, F-67000 Strasbourg, France\\
$^{5}$INAF - Osservatorio Astronomico di Cagliari, Via della Scienza 5, I-09047 Selargius (CA), Italy\\
}

\date{Accepted 2020 May 4. Received 2020 April 17; in original form 2020 February 14}

\pubyear{2020}

\begin{document}

\label{firstpage}
\pagerange{\pageref{firstpage}--\pageref{lastpage}}
\maketitle

\begin{abstract}

We present interferometric \hi\ observations of six double-disc stellar counterrotator (``2$\sigma$") galaxies from the \atlas\ sample.  Three are detected in \hi\ emission; two of these are new detections.
NGC~7710 shows a modestly asymmetric \hi\ disc, and the atomic gas in PGC~056772 is centrally peaked but too poorly resolved to identify the direction of rotation.
IC~0719, the most instructive system in this study, shows an extended, strongly warped disc of $\sim$ 43 kpc diameter, with a faint tail extending towards its neighbor IC~0718.  The gas has likely been accreted from this external source during an encounter whose geometry directed the gas into misaligned retrograde orbits (with respect to the primary stellar body of IC~0719). In the interior, where dynamical time-scales are shorter, the \hi\ has settled into the equatorial plane forming the retrograde secondary stellar disc. This is the first direct evidence that a double-disc stellar counterrotator could be formed through the accretion of retrograde gas.
However, the dominant formation pathway for the formation of $2\sigma$ galaxies is still unclear.   The \atlas\ sample shows some cases of the retrograde accretion scenario and also some cases in which a scenario based on an unusually well-aligned merger is more likely. 

\end{abstract}



\section{Introduction} \label{sec:intro}

Despite their relatively smooth and featureless appearances in shallow optical images, early-type galaxies (ellipticals and lenticulars, or slow- and fast-rotators) show an impressive diversity in their kinematic structures and frequent cases of internal misaligned substructures.
For example, the \atlas\ survey \citep{cappellari_a3d1} found that 4\% or 11 of 260 early-type galaxies actually consist of two dynamically cold, flattened stellar discs that are coplanar but counterrotating with respect to one another \citep{davor}.  
Beyond these double-disc stellar counterrotators, an additional 7\% of the \atlas\ sample (19 of 260) host multiple co-located stellar populations whose angular momenta are misaligned (the kinematically decoupled cores).  
These might be examples of a similar phenomenon to the double-disc stellar counterrotators.  A further 6\% (16 of 260) host two flattened discs whose specific angular momenta are aligned but they have different spatial distributions, so their velocity maps show local extrema both at small radii and again at larger radii, and these are prograde versions of the stellar counterrotators.
Thus, the double-disc stellar counterrotators may be considered the most extreme versions of a more general phenomenon that is evident in nearly 20\% of nearby early-type galaxies.

Furthermore,  35\% of the fast rotators have their ionized gas kinematically misaligned with respect to their stellar rotation by more than 40\arcdeg\ \citep{davis11,jin2016,ene2018,bryant2019}, and a similar misalignment fraction is observed in \hi\ \citep{serra14}.
Misaligned substructures are evidently common in early-type galaxies, and
a comprehensive picture of galaxy formation ought to be able to explain their incidence and
properties.

Our goal in the present paper is to understand the formation of the double-disc stellar counterrotators and to place them into the broader context of galaxy evolution.
In general, having multiple kinematic components suggests that a galaxy may have experienced multiple episodes of assembly, 
and in particular for the double-disc stellar counterrotators (also called ``two $\sigma$-peak" or ``2$\sigma$" galaxies), two classes of models have been discussed.
One model involves an ordinary disc galaxy that accretes a large quantity
of approximately retrograde cold gas, and after the gas has settled into the equatorial plane, it grows a retrograde stellar disc out of the accreted gas \citep[e.g.][]{vergani,coccato15,mitzkus}.
\citet{algorry} describe a variant in which a galaxy at the intersection of two cosmic filaments
could accrete gas first with one spin, from one dominant filament, and then with the opposite spin
(from another filament).
This scenario predicts that one of the stellar discs should be
younger than the other, and any cold gas that remains should rotate like the younger disc.

Another class of models involves a major binary merger of two galaxies in a rather special geometry --
the two galaxies and their mutual orbit are all nearly coplanar, and at least
one of the incoming galaxy spins is antiparallel to the pair's mutual orbital angular
momentum \citep[e.g.][]{bournaud05,crocker4550,bois}. 
Those simulations have shown that in this special aligned geometry, the stellar discs can survive
the merger more or less intact, with some dynamical heating.
The frequency of interactions with this special geometry is not well known.
The prototypical NGC~4550 is believed to have been formed this way, based on its CO kinematics \citep{crocker4550}.  The two stellar discs have different scale heights, and the CO rotates like the {\it thicker}
disc, which is the opposite to what one would expect for the accretion model (where the
younger disc should also be dynamically colder).
Understanding the relative importance of these two formation scenarios is necessary in order to check whether modern cosmological simulations of galaxy formation are correctly reproducing these galaxies.

As noted above in the case of NGC~4550, observations of the cold gas in double-disc stellar counterrotators can give crucial clues to their histories.
\hi\ is particularly valuable as it can often be traced to large radii and it is therefore more sensitive to interactions with other galaxies.  As the dynamical time-scales at large radii are long, \hi\ will retain those signatures of interactions for longer times than ionized or molecular gas tracers.  
Thus, we present \hi\ observations of six ``2$\sigma$" (stellar counterrotator) galaxies; three are detected, and one shows strongly disturbed kinematics that clearly point to the gas accretion formation scenario discussed above.
We discuss implications for the broader understanding of the formation of the 2$\sigma$ galaxies.

\section{Target selection}

The \atlas\ sample \citep{cappellari_a3d1} is a volume-limited sample of 260 early-type galaxies closer than about 40 Mpc, with stellar masses greater than 10$^{9.9}$ \solmass.  The heart of the survey is optical IFU spectroscopy, giving two-dimensional maps of the galaxies' stellar kinematics, stellar populations, ionized gas distribution and kinematics, and star formation history.  And while similar optical data exist for other (generally more distant) surveys, the \atlas\ sample is the best studied volume-limited sample at this time, so it provides the largest and most homogeneous selection of 2$\sigma$ galaxies for study.  It also has the most complete set of cold gas data, including observations of CO emission from all and \hi\ from most \citep{a3dco,serra12}.

The eleven 2$\sigma$ galaxies in \atlas\ \citep[Table \ref{tab:summary};][]{davor} are
identified using a combination of their stellar velocity fields and velocity dispersions.
The stellar discs have roughly comparable masses but slightly different scale heights and/or scale lengths, so they dominate the mean velocity field in different regions, and the sense of rotation in the velocity field switches at larger heights or radii.  
They also have unusually large off-nuclear peaks in the stellar velocity dispersion.  Those peaks are located along the major axis where the two counterrotating discs contribute roughly equally to the line-of-sight velocity distribution, which is where the sense of mean rotation changes.  This distinctive behavior in the velocity dispersions is responsible for the ``two $\sigma$-peak" name.   Their stellar kinematic maps can be seen in Figures C5 and C6 of \citet{davor}.

In addition to the eleven listed here, the \atlas\ member NGC~0524 has also been studied as a double-disc stellar counterrotator \citep{katkov11}.  The luminosity of its secondary disc is low, which makes it difficult to identify.  For the purposes of homogeneity we do not discuss NGC~0524 here, though we note that it does have a relaxed CO disc \citep{crocker11}, and it serves as a reminder that more sensitive data may identify additional 2$\sigma$ galaxies in the \atlas\ sample.

All eleven of the \atlas\ 2$\sigma$ galaxies were searched for CO emission \citep{a3dco} but not all of them were previously searched for \hi\ emission, due to the Declination limitations of the WSRT \citep{serra12}.  
\hi\ in NGC~4191 was discussed by \citet{young4191}.  The present paper fills in \hi\ observations of the remaining low-Dec galaxies not previously observed at WSRT, so that all eleven have sensitive CO searches and interferometric \hi\ searches.

\begin{table*}
\begin{centering}
\caption{Cold gas data for $2\sigma$ galaxies. \label{tab:summary}}
\begin{tabular}{lccclll}
\hline
Name &  log \mstar & log \mhtoo & log M(\hi) & \hi\ refs & Formation & Formation \\ 
       &   (\solmass) &   (\solmass)  &   (\solmass) &             &   & references\\
\hline
IC 0719   & 10.31 & 8.26 (0.04) & 9.03 (0.02) & * & gas accretion & \citet{katkov13}, \citet{pizzella} \\
NGC 0448  & 10.43 & $<$ 7.74    & $<$ 7.38 & * & merger & \citet{katkov16}, \citet{nedelchev}\\
NGC 3796  & 9.97   & $<$ 7.51  & $<$ 7.10 & S12 &  & \\
NGC 4191  & 10.47 & $<$ 7.94    & 9.57 (0.01) & Y18 & gas accretion & \citet{coccato15}, \citet{young4191}\\
NGC 4259  & 10.10 & $<$ 7.97  & $<$ 8.43 & * & & \\ 
NGC 4473  & 10.73 & $<$ 7.07  & $<$ 6.86 & S12 & & \\
NGC 4528   & 10.05 & $<$ 7.15 & $<$ 7.18 & S12 & & \\
NGC 4550   & 10.13 & 6.91 (0.08)  &  $<$ 6.89 & S12 & merger & \citet{crocker4550}, \citet{coccato13}, \citet{johnston}\\ 
NGC 4803   & 10.14 & $<$ 7.98  & $<$ 7.57 & * &  & \\
NGC 7710   & 10.02 & $<$ 7.80   &  8.74 (0.02) & * & gas accretion? & * \\
PGC 056772  & 10.05 & 8.19 (0.05)  & 8.11 (0.06) & * & gas accretion?  & *\\
\hline
\end{tabular}
\end{centering}

Stellar masses are from \citep{cap:a3dJAM}.  \htoo\ masses are all from \citet{a3dco}, except NGC~4550 from \citet{crocker4550}.  \\
\hi\ references: * = This paper; S12 = \citet{serra12}; Y18 = \citet{young4191}.
\end{table*}

\section{Observations}

IC~0719, NGC~0448, NGC~4803, NGC~7710, and PGC~056772 were observed with the Karl G.\ Jansky Very Large Array (VLA)
in September 2018, in project 18A-226, for a total of two hours each (and four hours on NGC~7710).  These data were obtained in the D configuration, giving baselines from 0.2 to 5
k$\lambda$ and resolutions on the order of an arcminute.  
The native velocity resolution of the data is 3.3 \kms.
The flux, bandpass, and time-dependent gain calibrations were made following standard
procedures, using 3C286 or 3C48
once per observing session as the primary flux
and bandpass calibrator, and secondary gain calibrators were
observed every 20 minutes.  
Calibration was performed using the eVLA scripted pipeline the CASA software package, version 5.3.0. 
The bootstrapped flux densities of the secondary gain calibrators are in good agreement
with previous measurements in the VLA calibrator database and the NRAO VLA Sky Survey \citep{nvss}.
An initial round of imaging revealed any bright line emission in the data, and the velocities
of that emission combined with the known target velocities were used to select frequency 
ranges for estimating continuum levels.
Continuum emission was subtracted using order 0 or 1 fits to the individual visibilities.
We produce final data cubes using the `natural' weighting scheme for maximum sensitivity.  If line emission is present, we also produce cubes of several different channel widths using the Briggs robustness technique and some tapering in the $uv$ plane to achieve better resolution and a rounder beam.  Cubes are cleaned to a residual level equal to the rms noise level. 

In addition, NGC~4259 was observed in project AT259 in 2001, in the VLA's C configuration.  General observing strategies, data reduction, and imaging are similar to those used for the newer data.  The data give baselines from 0.3 to 16 k$\lambda$ but they have only 30 minutes on source, so they are higher resolution and lower sensitivity than the other data discussed in this paper.  Furthermore, NGC~4259 is 8.7\arcmin\ away from the pointing centre, but the offset is not a serious problem as the half-power radius of the VLA is 15\arcmin\ and the antenna gain at the location of NGC~4259 is still a relatively high value of 0.82.  Thus these C configuration observations are not ideal for very faint, diffuse emission, but they can rule out the presence of bright \hi\ emission like that in IC~0719 or NGC~7710.  

The sensitivity, final imaged velocity resolution and velocity coverage of these \hi\ observations are indicated 
in Table~\ref{tab:images}.
Following the discussion in \citetalias{serra12}, which uses data of similar sensitivity and
resolution, we indicate the \hi\ column density sensitivity as a 5$\sigma$ signal in one 
channel of width 16.5 \kms. 
\citetalias{serra12} also make a careful analysis of the typical angular sizes
and velocity widths of detected \hi\ features in the \atlas\ survey.  Based on this
analysis, they adopt M(\hi) upper limits for the \hi\ nondetections
by calculating three times the statistical uncertainty in a sum over a data volume of 50 \kms\ and six synthesized beam areas 
(typically 1.2\e{4} square arcseconds or 110 kpc$^2$ to 450 kpc$^2$ at the distances of this sample).
That procedure is also adopted here, using the appropriate number of beams to make the solid angle of 1.2\e{4} square arcseconds.

\begin{table*}
\centering
\caption{\hi\ image properties for $2\sigma$ galaxies.\label{tab:images}}
\begin{tabular}{llccccccr}
\hline
Name & Distance & Vel.\ range & beam & beam & rms & N(\hi) lim & S(\hi) & M(\hi) \\ 
       & (Mpc) & (\kms)     & (\arcsec) & (kpc) & (mJy/bm) & ($10^{19}$\persqcm) & (\jykms) & (\solmass) \\
\hline
 IC~0719  &  29.4  & 288 -- 3376   &   111 $\times$ 57 &  15.8 $\times$ 8.1 & 0.85 & 1.23 &  5.30 $\pm$ 0.27 & (1.08$\pm$0.06)\e{9}  \\ 
                 &           &                       &  82 $\times$ 65 &  11.7 $\times$ 9.3 & 1.20  &  1.21 &                             &        \\
 NGC~0448  & 29.5 & 366 -- 3438 & 111 $\times$ 55 & 15.9 $\times$ 7.8 & 0.98 & 1.46 & $<$ 0.12 & $<$ 2.4\e{7} \\
 NGC~4803  & 39.4 & 1071 -- 4226 & 109 $\times$ 50 & 20.8 $\times$ 9.5 & 0.83 & 1.39 & $<$ 0.10 & $<$ 3.7\e{7} \\
 NGC~7710  & 34.0 & 834 -- 3898 & 68 $\times$ 56 & 11.3 $\times$ 9.2 & 0.59 & 1.41 & 2.00 $\pm$ 0.09 & (5.45$\pm$0.24)\e{8} \\
 PGC~056772  & 39.5 & 1120  -- 4154 & 76 $\times$ 58 & 14.5 $\times$ 11.1 & 0.66 & 1.36 & 0.35 $\pm$ 0.05 & (1.30$\pm$0.19)\e{8} \\
 NGC~4259 & 37.2 & 1803 -- 3080 & 19.6 $\times$ 15.7 & 3.5 $\times$ 2.8 & 1.47 & 54 & $<$ 0.83 & $<$ 2.7\e{8} \\ 
\hline
\end{tabular}

Distances are taken from \citet{cappellari_a3d1}.  The velocity
range in column 3 indicates the usable range covered by the data. The noise (rms) is quoted in a channel 16.5 \kms\ wide, except for NGC~4259, where 20.6 \kms\ is the native resolution of the data, and for the higher resolution image of IC~0719, which was made with 9.9 \kms\ channels.
The column density limit represents 5$\sigma$ in one channel. The integrated flux density limit for nondetections represents three times the statistical uncertainty in a sum over 
50 \kms\ and 1.2\e{4} square arcseconds, as in \citetalias{serra12}.  For the new D configuration data that solid angle corresponds to $\sim$ 2 beams and for the C configuration data on NGC~4259 it is 34 beams.
\end{table*}

\section{Results}

We find \hi\ emission in NGC~7710, PGC~056772, and IC~0719, but not in NGC~0448, NGC~4803, or NGC~4259 (Table \ref{tab:images}).  These results bring the
\hi\ detection rate in the \atlas\  2$\sigma$ galaxies to 4/11, comparable to the rate in \atlas\  as a whole \citepalias{serra12}.  Here we note the nondetections briefly and describe the \hi\ emission in NGC~7710, PGC~056772, and IC~0719.

\subsection{Nondetections -- NGC~4803, NGC~0448, NGC~4259}

No \hi\ emission is found towards NGC~4803.  The VLA data for this field show some \hi\ emission from UGC~08045 and possibly also from the vicinity of NGC~4795, but they are far away and there's no sign of any emission towards NGC~4803 or between it and the other detections; more information is presented in the Appendix.  No significant \hi\ emission from any source is detected in the field of NGC~0448.
And as noted in the previous section, 
the observations we are using for NGC~4259 were pointed at the spiral NGC~4273, 8.7\arcmin\ away (94 kpc in projection); NGC~4273 is strongly detected and is shown in the Appendix, but NGC~4259 is not detected.

\subsection{NGC 7710 and PGC 056772}

\begin{figure}
\includegraphics[scale=0.5, trim=2cm 7.5cm 3cm 5cm,clip]{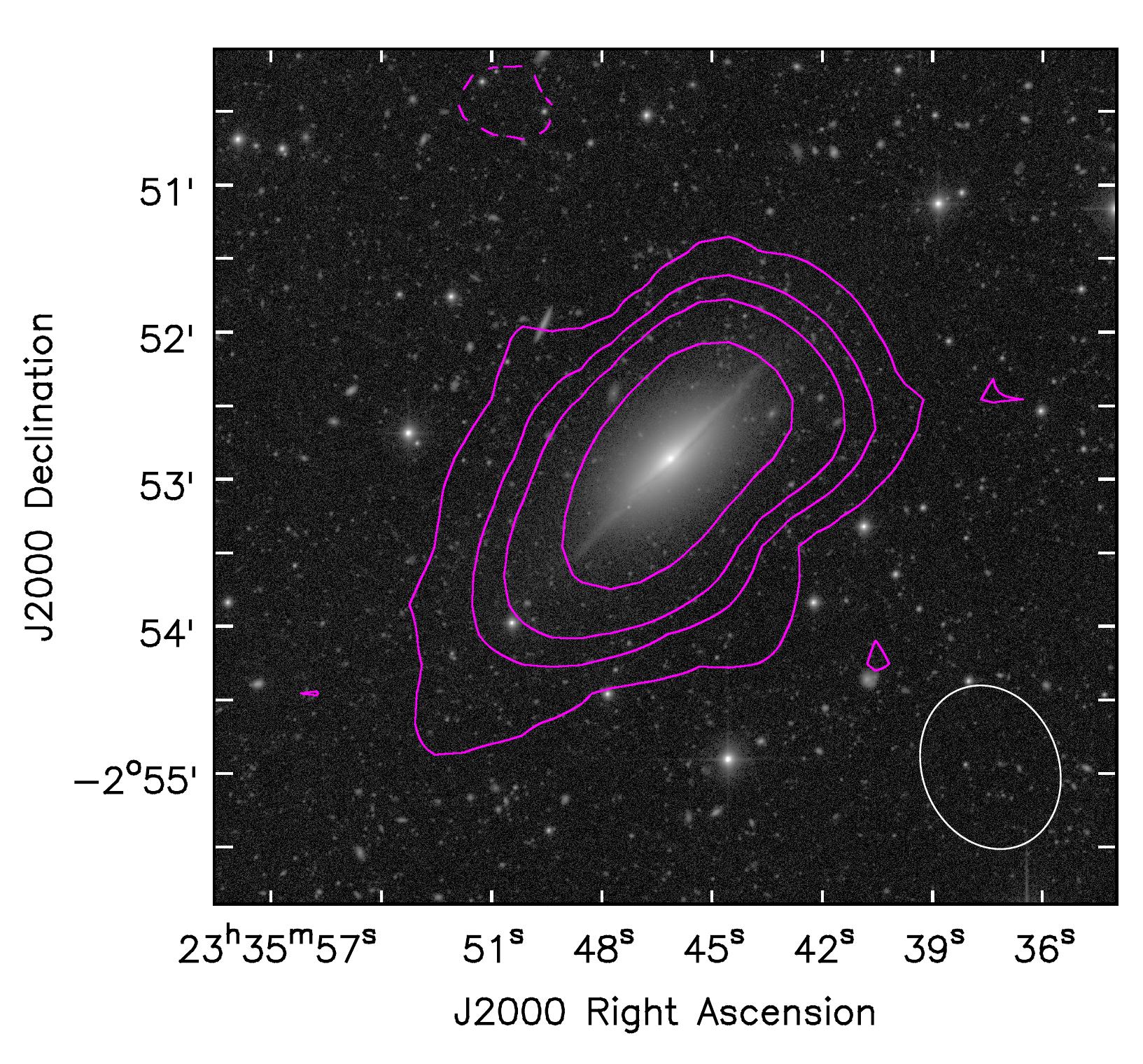}
\caption{\hi\ integrated intensity in NGC~7710.  The grayscale is the MATLAS $g$ image from \citet{duc15}, and the contours are the \hi\ column density at levels of $(-1, 1, 3, 5, 10, 30)\times 1.41\times 10^{19}$\persqcm, where 1.41\e{19}\persqcm\ is the nominal sensitivity in these data (Table \ref{tab:images}). The resolution of the \hi\ data is indicated by the ellipse in the lower right corner.\label{7710m0}}
\end{figure}

\begin{figure}
\includegraphics[scale=0.65,clip]{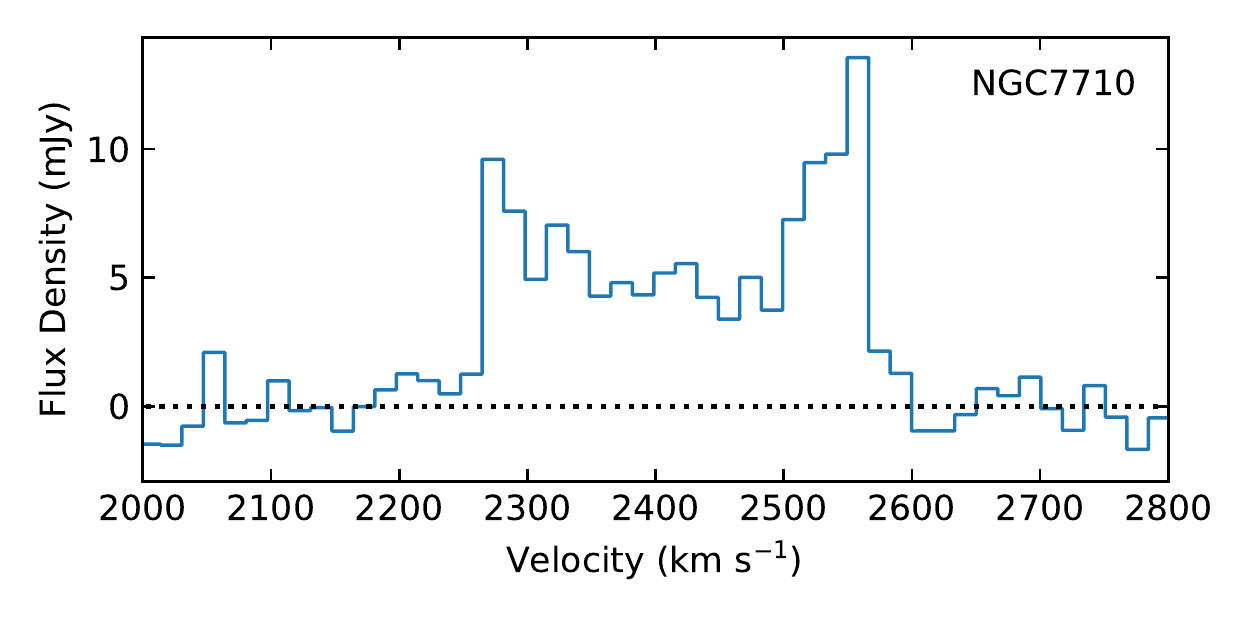}
\caption{\hi\ spectrum of NGC~7710.\label{7710spec}}
\end{figure}

\begin{figure}
\includegraphics[width=\columnwidth, trim=0.5cm 0.5cm 0.5cm 1cm, clip]{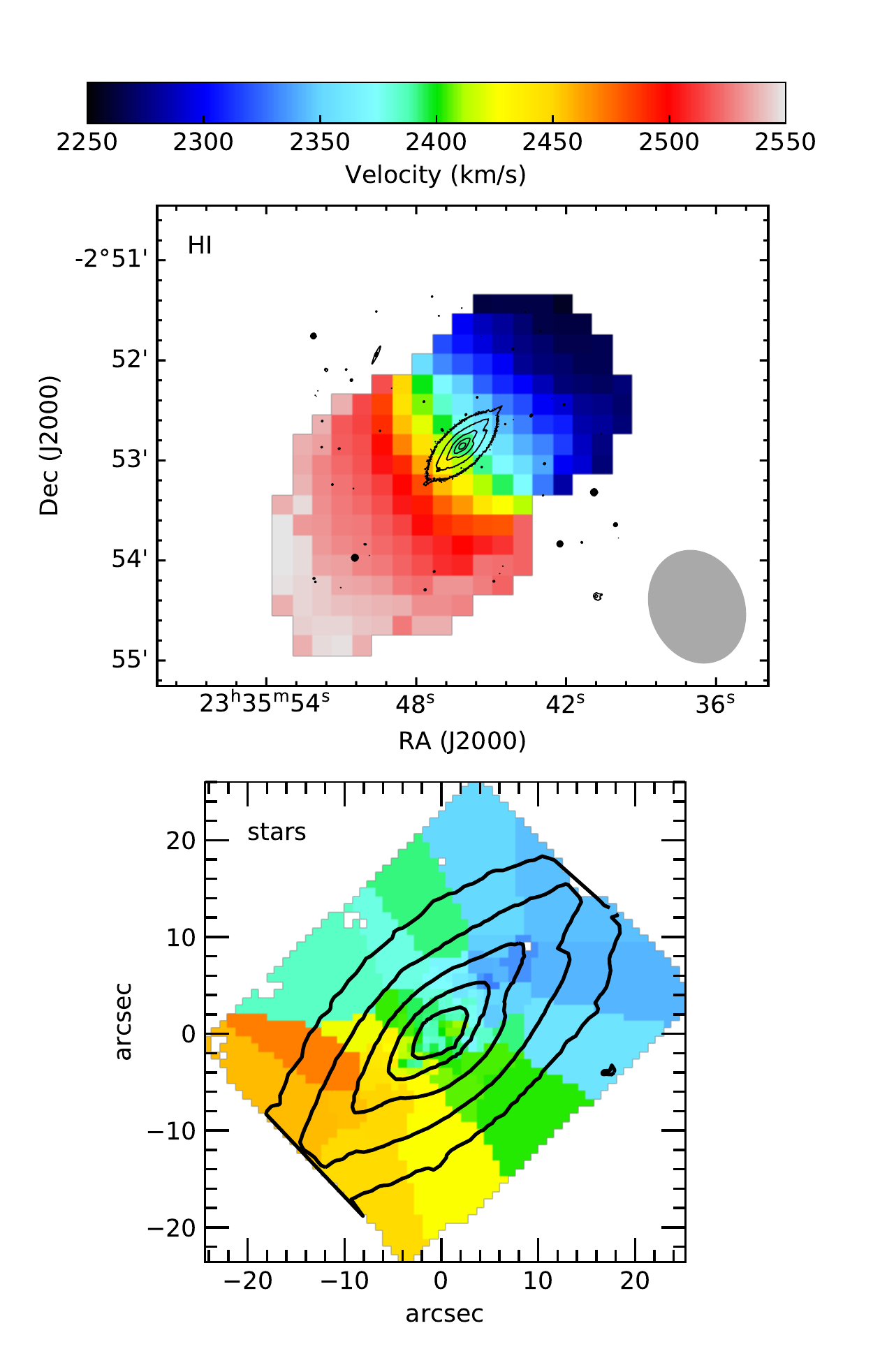}
\caption{\hi\ and stellar velocity fields of NGC~7710.  In both panels, the black contours show the stellar isophotes.  In the stellar velocity field, the counterrotating feature is visible inside the brightest contour, though it is more prominent in the stellar velocity dispersion data \citep{davor}.\label{7710vels}}
\end{figure}

NGC~7710 is unusual amongst the 2$\sigma$ galaxies in that its optical image (Figure \ref{7710m0}) clearly shows multiple components: a thick spheroid and a very thin, nearly edge-on disc with an associated dust lane.
\hi\ emission in NGC~7710 shows the classic double-horned profile typical of galactic discs (Figure \ref{7710spec}).
The gas distribution and kinematics (Figures \ref{7710m0} and \ref{7710vels}) show little sign of recent disturbance, though there is a mild asymmetry with approximately 30\% more \hi\ on the receding (southeast) side of the galaxy than the approaching side, and the gas is more extended on that side as well.  It is important to note that the \hi\ distribution is not resolved in the direction of the minor axis, so the distribution by itself offers no clues about whether the \hi\ is associated with the thicker stellar spheroid or the thinner disc.  We assume it is associated with the thinner disc, as that's where the dust and the blue stellar populations are found.
The \hi\ rotates in the
same direction as the ionized gas.  That is also the same as the sense of rotation in the stellar velocity field at 
radii $\gtrsim$ 4\arcsec\ \citep[Figure \ref{7710vels};][]{davor}.

\begin{figure}
\includegraphics[scale=0.5, trim=2cm 7.5cm 3cm 5cm,clip]{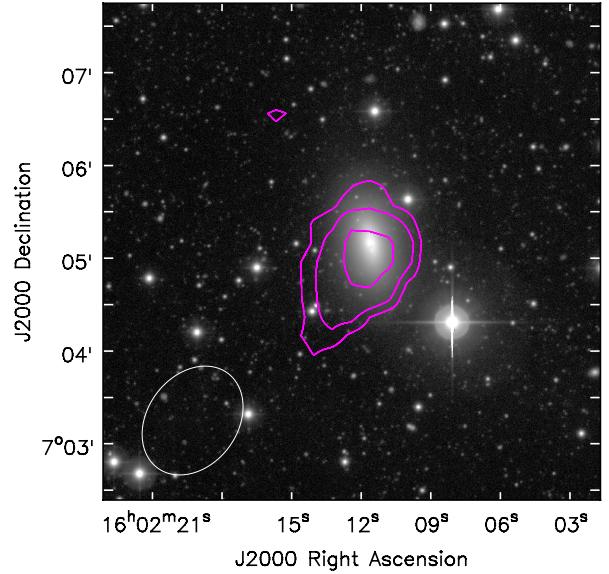}
\caption{\hi\ integrated intensity in PGC~056772.  The grayscale is the MATLAS $g$ image \citep{duc15}, and the contours are the \hi\ column density at levels of $(1, 2, 4)\times 1.36\times 10^{19}$\persqcm, which is the nominal sensitivity in these data. The resolution of the \hi\ data is indicated by the ellipse in the lower left corner. \label{pgc056m0}}
\end{figure}

\begin{figure}
\includegraphics[scale=0.65,clip]{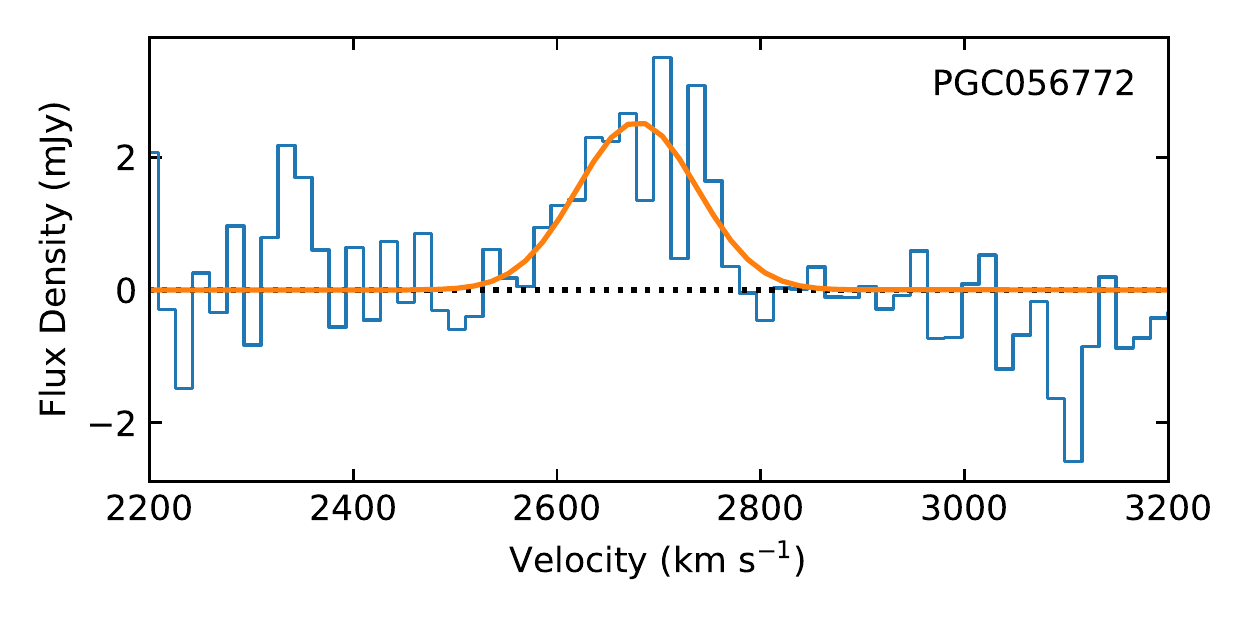}
\caption{\hi\ spectrum of PGC~056772, with a Gaussian fit overlaid.\label{pgc056spec}}
\end{figure}

\begin{figure}
\includegraphics[width=\columnwidth]{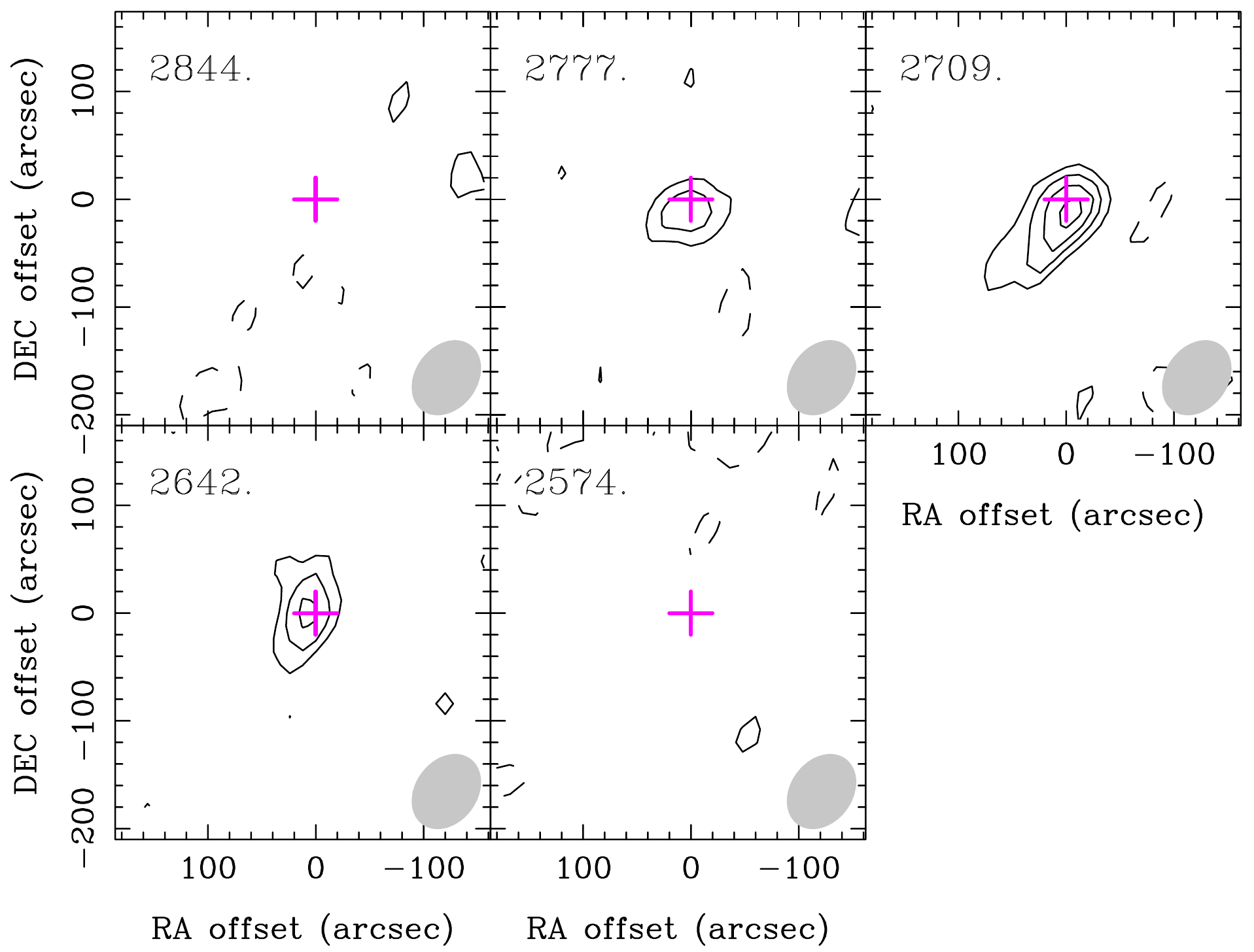}
\caption{\hi\ channel maps of PGC~056772.  The ellipse illustrates the beam size; the magenta cross marks the optical centre of the galaxy, and the velocity of each channel is indicated in the top left corner.  Contour levels are $(-2, 2, 3, 4, 5)$ times the rms noise in this image cube (0.37 \mjb).}\label{pgc056chans}
\end{figure}

PGC~056772, like most of the 2$\sigma$ galaxies, has a rather unremarkable optical image (Figure \ref{pgc056m0}) and its multiple stellar components are much more obvious in the stellar kinematics \citep{davor} than in the optical surface brightness distribution.  It also shows very faint \hi\ emission  (Figures \ref{pgc056m0} and \ref{pgc056spec}).  The galaxy has a 1.4 GHz continuum source with a flux density of 4.7 $\pm$ 0.4 mJy, so the coincidence of the line source and the continuum source boosts the reliability of the line detection.  The \hi\ emission is not well resolved, so we have little kinematic information (Figure \ref{pgc056chans}), and it is not yet possible to associate the \hi\ emission with a specific stellar component.  There is a hint of an extraplanar \hi\ tail, stretching to the southeast, but it should be verified with data of higher sensitivity and resolution.  Three dwarf galaxies in the field of PGC~056772 are also detected in \hi\ emission and identified in the catalog of \citet{habas2019}, and they are discussed in the Appendix.

\subsection{IC 0719}

IC~0719 is relatively dusty and blue for an early-type galaxy \citep[e.g.][]{davor,a3dcocarma}.  Its two stellar components are more obvious in the stellar velocity field than in the optical image, as is common, but in hindsight the stellar components can also be decomposed in an optical image \citep{katkov13,pizzella}.  IC~0719 is in a close pair (at 10\arcmin\ or 86 kpc in projection) with IC~0718, which has a similar velocity and a factor of 10 smaller stellar mass.  IC~0718 is not a member of the \atlas\ sample and is not known to be a 2$\sigma$ galaxy, but it is detected in \hi\ emission in our data and we discuss it in the context of a probable interaction with IC~0719.

\begin{figure}
\includegraphics[width=\columnwidth]{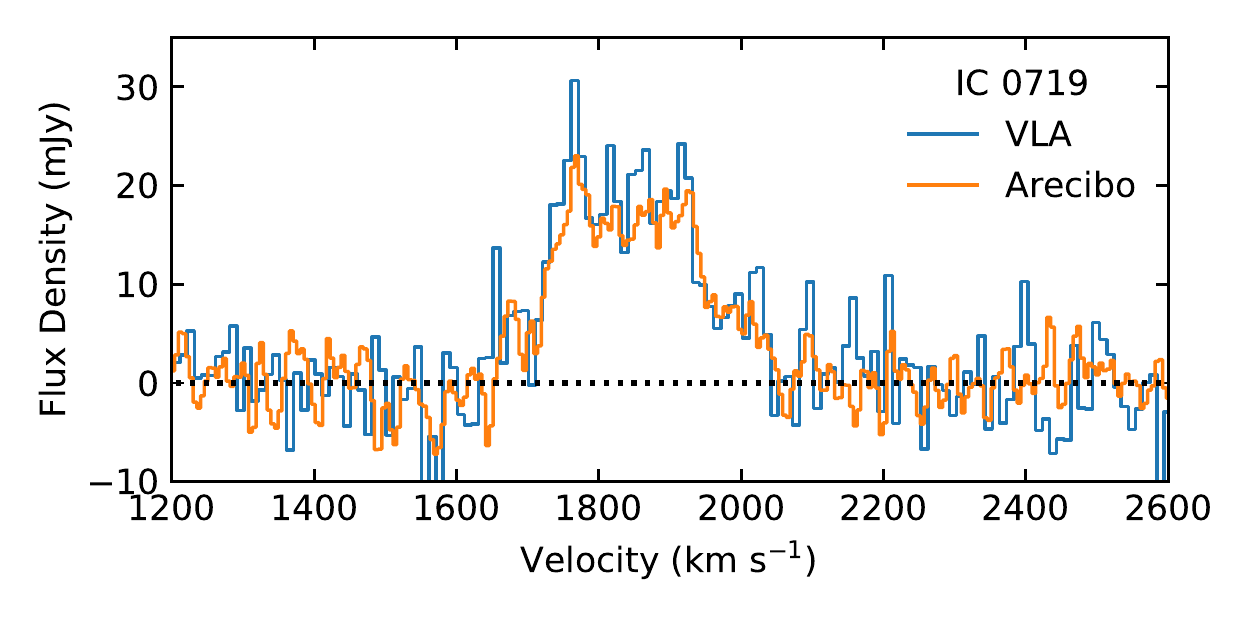}
\includegraphics[width=\columnwidth]{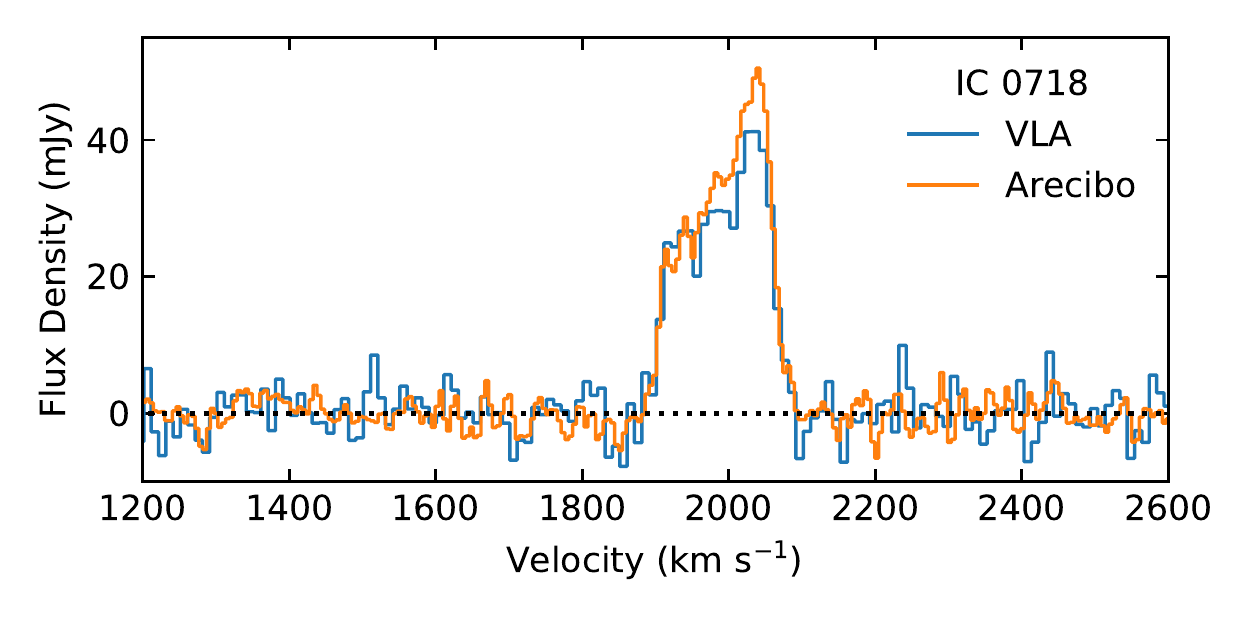}
\caption{Integrated \hi\ spectra of IC0719 and IC0718 from the VLA data and from Arecibo \citep{alfalfa}. \label{spec}
}
\end{figure}

The total flux that we recover in IC~0719 is 5.30 $\pm$ 0.27 \jykms, or (1.08 $\pm$ 0.06)\e{9} \solmass\ of \hi, which is consistent with the flux measured by Arecibo (Figure \ref{spec}).  To be specific, the flux recovered here is slightly larger than the values of 4.65 $\pm$ 0.12 \jykms\  and 4.08 $\pm$ 0.10 \jykms\ quoted by \citet{alfalfa} and \citet{grossi}, most likely because the emission is somewhat more extended than the Arecibo beam.  
In IC~0718 we recover 4.89 $\pm$ 0.16 \jykms, which is 13\% smaller than the Arecibo measurement of 5.53 $\pm$ 0.08.  The difference between those measurements may be attributed to the fact that IC~0718 is 10\arcmin\ from the VLA's field centre, where the primary beam gain is 0.72.  We have applied the primary beam correction, but the imaging quality is always degraded at large distances from the field centre, and extended smooth emission may be filtered out.

\begin{figure}
\includegraphics[width=\columnwidth,clip]{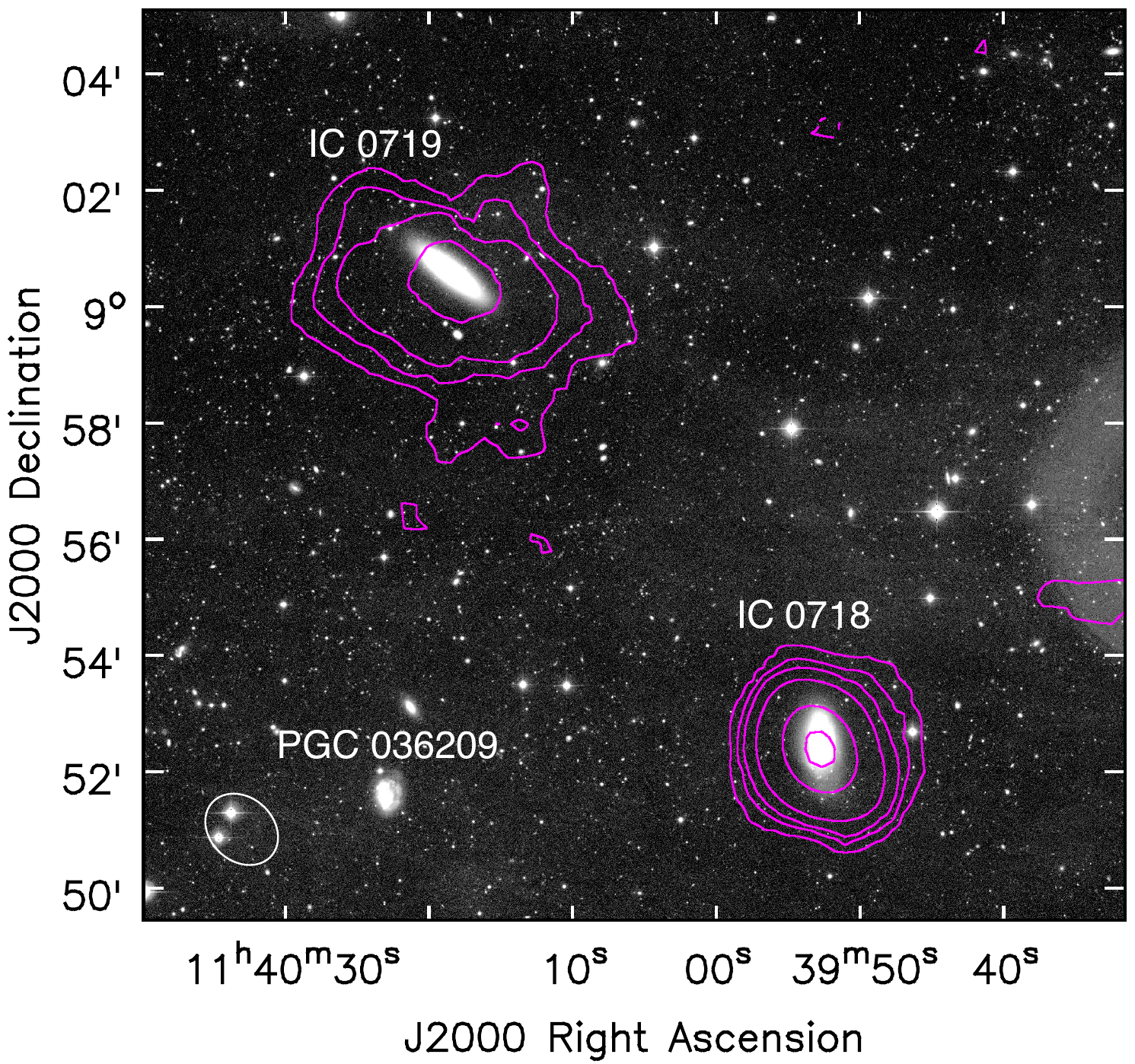}
\caption{\hi\ in IC~0719 and IC~0718.  The optical image is the MATLAS $r$ image \citep{duc15}.  Contours are $(-1, 1, 3, 5, 10, 30, 50)\times 1.2\times 10^{19}$\persqcm, where 1.2\e{19}\persqcm\ is the nominal sensitivity in this image (Table \ref{tab:images}).
PGC~036209 has a much higher velocity of 6372 \kms\ \citep{alfalfa} so it is a distant background object.
\label{both}
}
\end{figure}

\subsubsection{Disturbed \hi\ kinematics in IC~0719}\label{sec:hikin}

\begin{figure}
\includegraphics[width=\columnwidth, trim=5mm 0mm 0mm 0mm,clip]{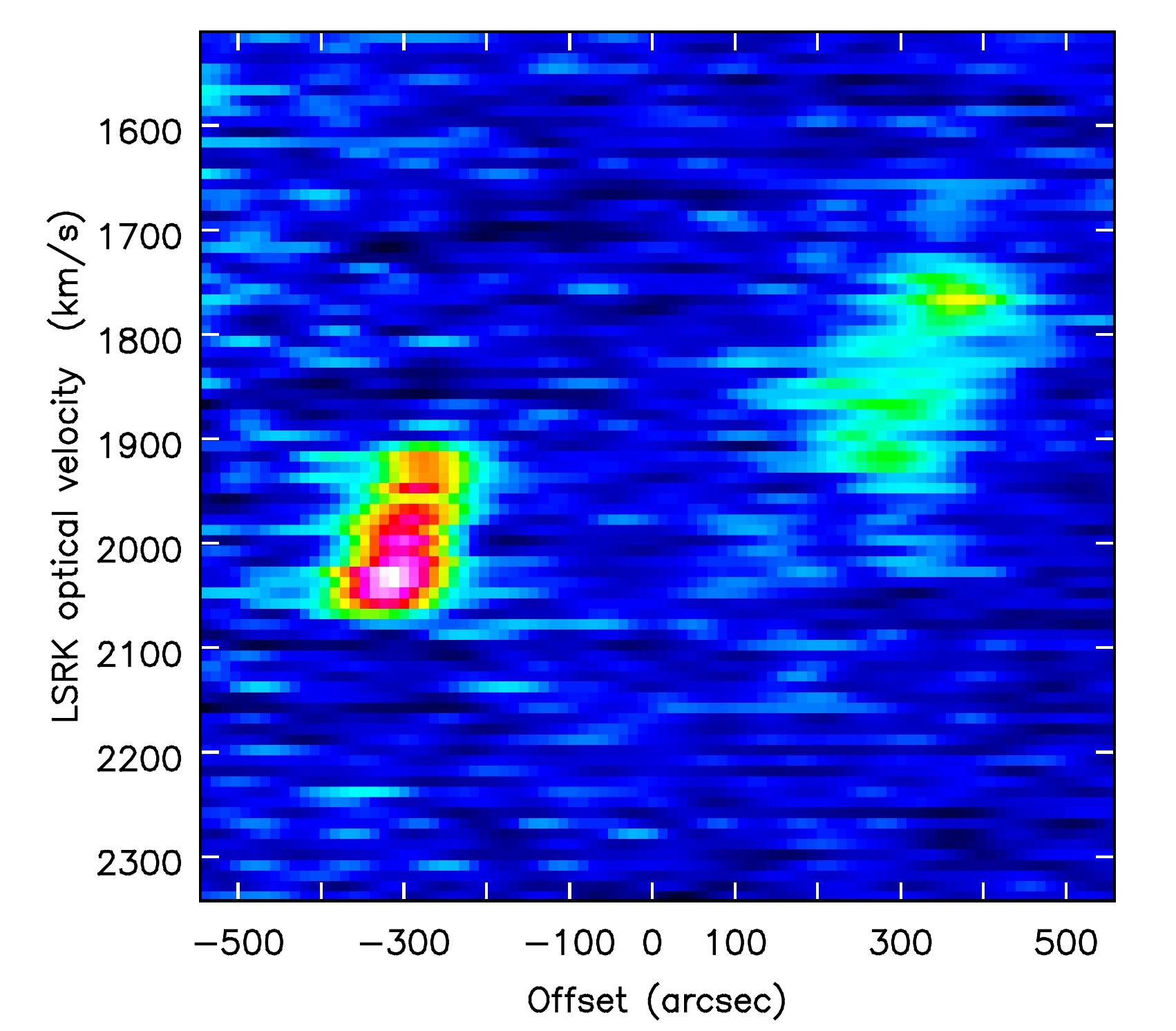}
\caption{Position-velocity slice between IC~0718 (left) and IC~0719 (right).  In this slice the emission is averaged over a region 204\arcsec\ wide ($\sim$ 2.5 beams). \label{pvslice}}
\end{figure}

IC~0719 has an irregular \hi\ distribution roughly 5\arcmin\ (43 kpc) in diameter in the east-west direction and 4\arcmin\ north-south (Figure \ref{both}).  A tail of faint \hi\ emission at column density $\sim$ 2\e{19}\persqcm\ extends at least 4\arcmin\ (34 kpc) from the centre of IC~0719 in the direction of IC~0718.  This tail is probably the same feature as the bridge shown between IC~0719 and IC~0718 in the Arecibo data of \citet{grossi}, and though those authors show the bridge to be complete, with no gaps, the VLA data do not show a full bridge (Figure \ref{pvslice}).   Grossi et al do not specify the peak \hi\ column density in the bridge, indicating only that it is greater than 6\e{18}\persqcm\ in the $\sim$ 4\arcmin\ Arecibo beam, and the nominal sensitivity in the present VLA data is 1.2\e{19}\persqcm.  Therefore the two datasets are not in direct conflict, especially if the emission is spread over 50 \kms\ or more.  It is also worth noting that single dish images are usually not cleaned to account for the presence of sidelobes of the beam, and Grossi et al do not mention cleaning, so there is always the possibility that the column densities in the bridge between two bright emission sources could be enhanced by an unfortunate sidelobe.  Deeper interferometric images would be the best way to probe the existence of a complete bridge in more detail.

\begin{figure*}
\includegraphics[width=\textwidth, trim=0.5cm 0cm 1cm 3cm, clip]{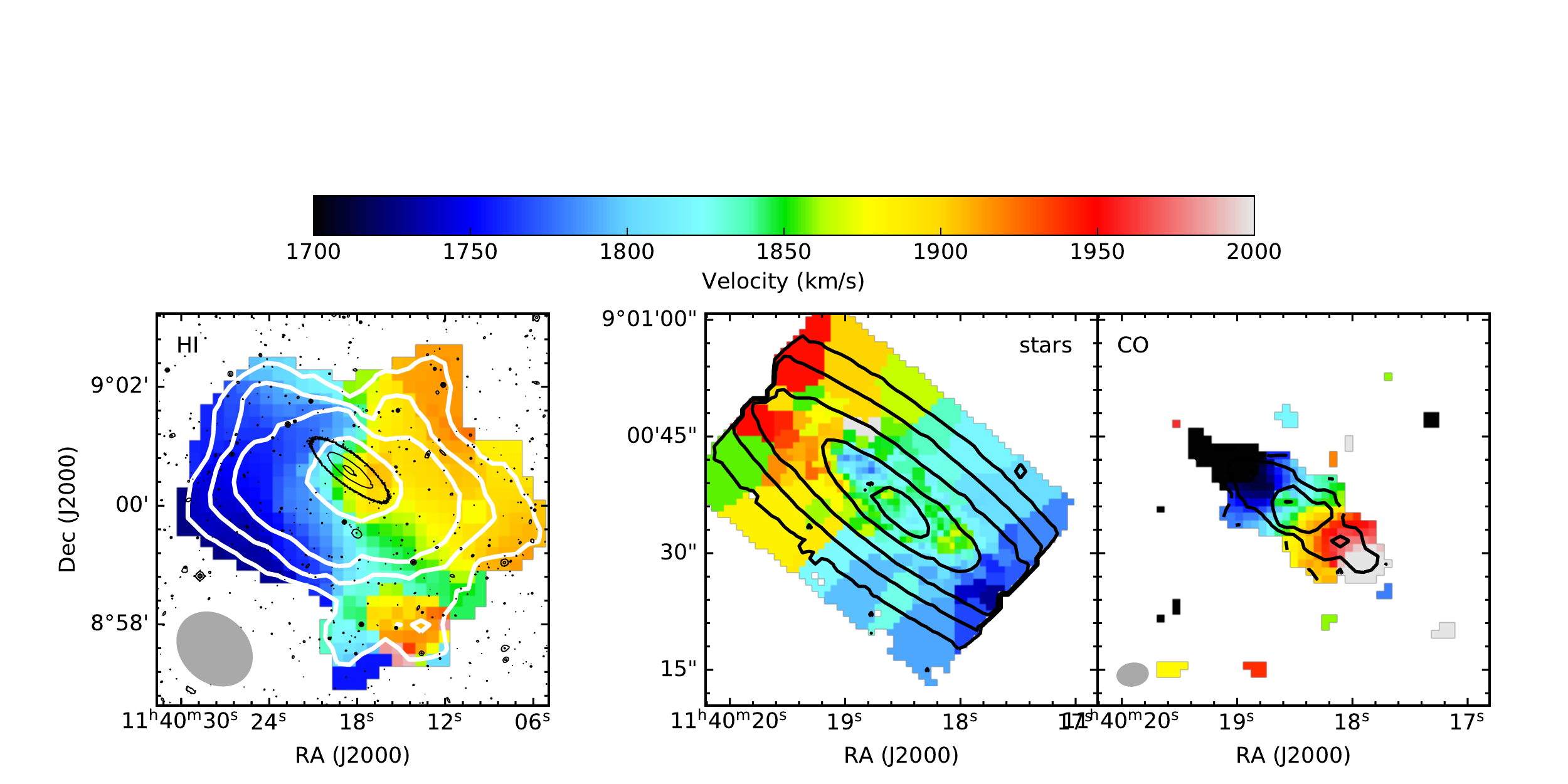}
\caption{\hi, stellar and CO velocity fields of IC~0719.  For \hi, the white contours show the column density as in Figure \ref{both} and the black contours are from the optical image to show where the main stellar body of the galaxy is located.  In the stellar velocities (centre panel), the black contours are again the stellar surface brightness.  The primary stellar component defined by \citet{pizzella} is receding on the northeast side of the galaxy; the secondary component is receding on the southwest side of the galaxy.  The secondary component dominates the stellar velocity measurements in the annulus between the highest contour and the second-highest contour (radii 4\arcsec\ to 10\arcsec); the primary component dominates both at smaller radii and at larger radii.  In the CO field, black contours show the integrated CO surface brightness.  Beam sizes are indicated by the ellipses.}
\label{ic0719allvelfields}
\end{figure*}

\begin{figure*}
\includegraphics[scale=0.9, trim=5mm 5mm 0mm 5mm]{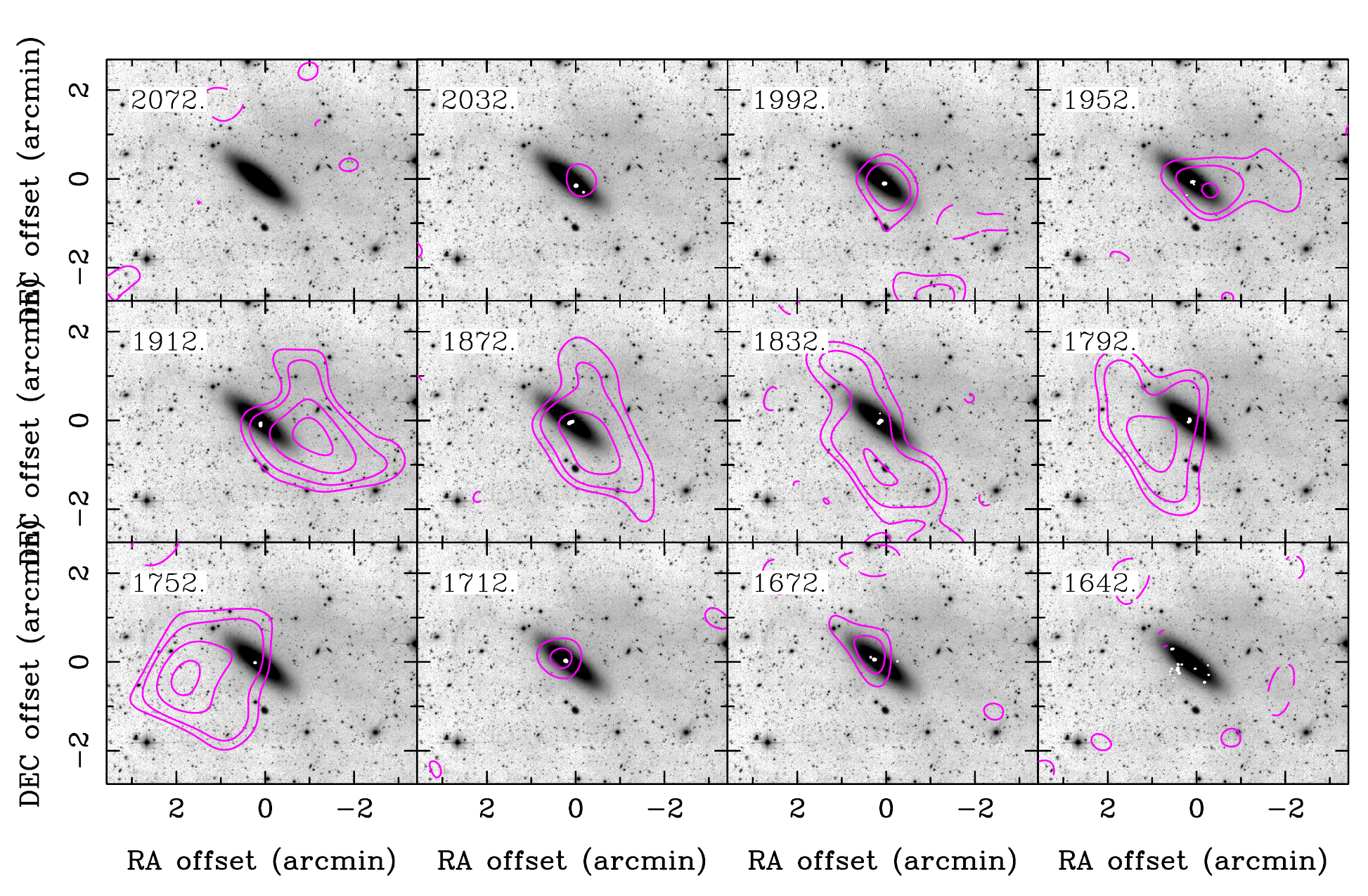}
\caption{Channel maps of IC~0719.  The greyscale background is the MATLAS $g$ image; pink contours are \hi\ at levels of ($-3$, $-2$, 2, 3, 5, 7) $\times$0.85 \mjb.
The velocity of each channel is written in the top left corner.  CO emission in the interior of the galaxy \citep{a3dcocarma} is indicated in tiny white contours near the galaxy centre.
\label{chans}}
\end{figure*}

The velocity fields and channel maps (Figures \ref{ic0719allvelfields} and \ref{chans}) show that in the inner body of the galaxy, the \hi\ kinematics are aligned with the photometric major axis of the galaxy,  which also aligns with the ionized gas and CO kinematics.   
The \hi\ kinematic position angle at small radii is 243\arcdeg $\pm$ 3\arcdeg, compared to 229\arcdeg $\pm$ 3\arcdeg\ for the molecular gas and 231.9\arcdeg $\pm$ 0.1\arcdeg\ for the stellar photometric major axis \citep{davis11,davor}.
The sense of rotation of the \hi\ matches that of the smaller, secondary stellar disc but is retrograde with respect to the bulk of the stars \citep{katkov13,pizzella}.
And though the mismatch between the \hi\ and CO resolution is great, the channel maps show that the highest velocity \hi\ emission in the galaxy forms a compact disc coincident with and co-rotating with the CO; the feature is most easily visible in the channel maps on the receding side at 2032 and 1992 \kms\ and on the approaching side at 1712 and 1672 \kms.

However, at radii $>$ 1\arcmin, the kinematic position angle of the \hi\ rotates to be 288\arcdeg $\pm$ 7\arcdeg.  This outer kinematic position angle is strongly misaligned, as it is 124\arcdeg\ ($\pm$ 7\arcdeg) away from the kinematic position angle of the bulk of the stars.
The large-scale \hi\ disc is irregular and blobby, with a lower inclination than the inner \hi\ and CO disc; it is visible in the channel maps between 1952 \kms\ and 1752 \kms.
And since the resolution of the \hi\ data is rather low, the radius of the warp in the \hi\ kinematics is rather poorly constrained, but  we note that the CO kinematics are quite regular.  We infer that the warp occurs somewhere between the outer edge of the CO disc (radius 15\arcsec\ = 2.1 kpc) and 1\arcmin.

\subsubsection{IC~0718 as the source of the gas in IC~0719}\label{sec:0718}

\begin{figure}
\includegraphics[width=\columnwidth, trim=2cm 0cm 3cm 0.5cm,clip]{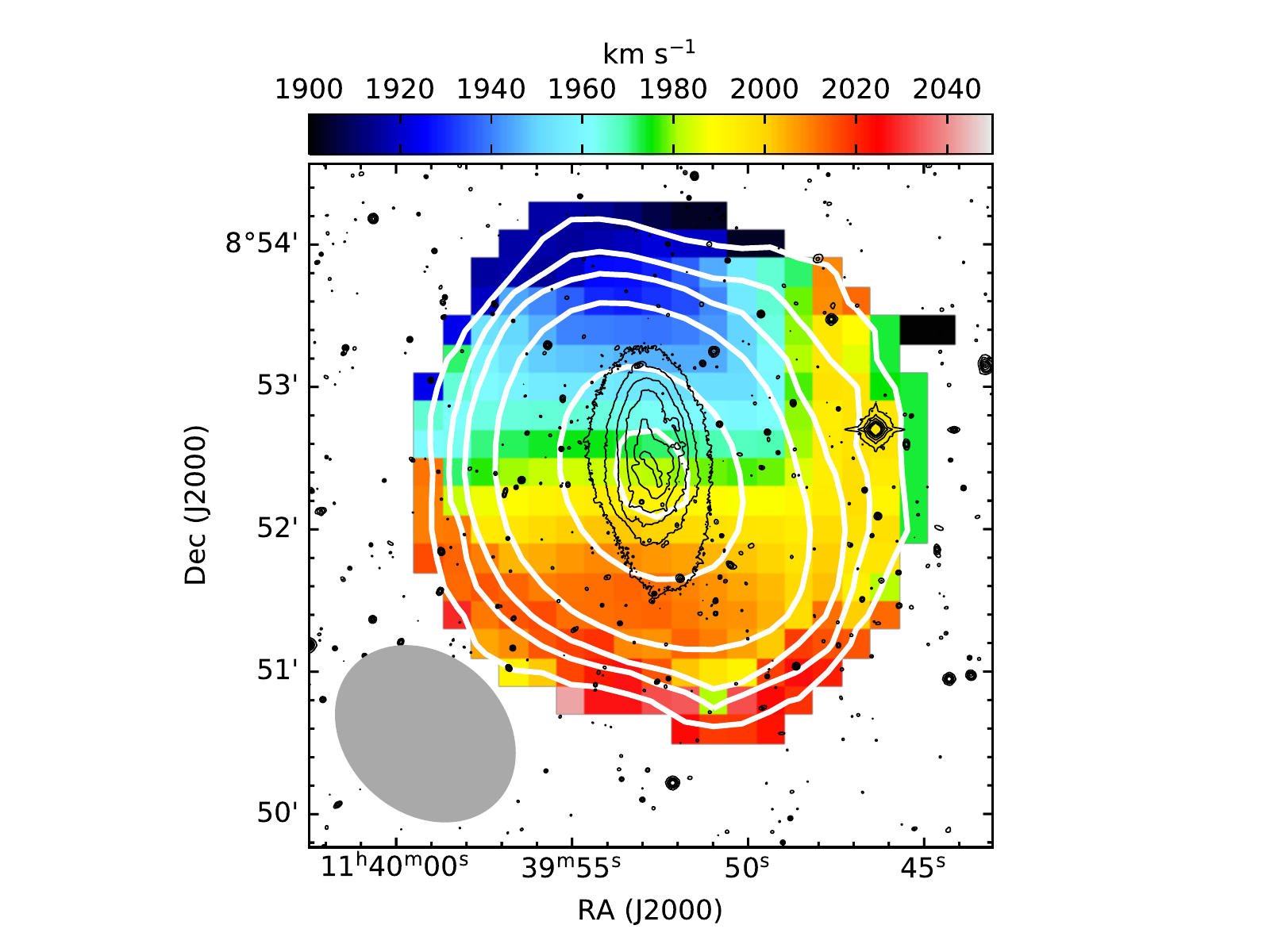}
\caption{Velocity field of IC~0718 (neighbor to IC~0719).  Thick white contours are the \hi\ column density, as in Figure \ref{both}, and thin black contours are from the MATLAS $g$ image.  The beam size is indicated by the gray ellipse.\label{ic0718}}
\end{figure}

IC~0718, the companion, is a late-type galaxy and it also shows signs of an interaction.  Its stellar distribution is asymmetric, with a relatively bright arm extending to the north and a stubby, faint tail extending to the south (Figures \ref{both} and \ref{ic0718}).  The \hi\ spectrum is also strongly asymmetric (Figure \ref{spec}), suggesting the \hi\ distribution is far from a relaxed equilibrium state; the brighter \hi\ emission is on the south side of the galaxy, coinciding with the low surface brightness stellar tail. The velocity field (Figure \ref{ic0718}) is mostly regular but that is probably a consequence of the relatively poor angular resolution in the \hi\ data.  
These stellar and \hi\ asymmetries demonstrate  that the galaxy has been involved in a recent interaction and therefore it is most likely the source of the \hi\ recently accreted onto IC~0719.

An hypothesized gas transfer from IC~0718 to IC~0719 suggests that IC~0718 should be \hi-deficient for its stellar properties.
The stellar mass of IC~0718 \citep{leroy2019} gives $\mathrm{M_{\hi}/M_\star} = 0.58$, and according to the analysis of \citet{brown2015} this value is about a factor of 2.6 low for a galaxy of IC~0718's stellar mass and NUV-r color.  By this measure IC~0718 is unusually \hi-deficient, and its expected \hi\ content could easily supply the gas we now detect in IC~0719.

Estimates of the \hi\ size of IC~0718 give less conclusive answers to the question of whether its \hi\ disc is truncated.  From the centroids of the emission in the extreme channels we estimate an \hi\ diameter of 10.6 kpc (we assume the same distance as for IC~0719).  In contrast, a model fit to the \hi\ image cube using the 3DBarolo code \citep{barolo} suggests that the \hi\ diameter at a surface density of 1 \solmass~pc$^{-2}$ is 19 kpc, and this latter value is entirely consistent with the \hi\ mass-size relation of \citet{wang2016}.  The extremely poor resolution in these \hi\ data means that comparisons to the mass-size relation are probably less reliable than comparisons to the stellar mass. 

For completeness we should also consider the possibility that the interaction between the two galaxies operates in the other direction, i.e.\ that IC~0719 had at least $10^9$\solmass\ of \hi\ before its encounter, and the disturbed  \hi\ kinematics might have arisen not because the gas has recently fallen in but because the gas has recently been pulled out by tidal forces.  That scenario is possible, but it is not the simplest scenario to explain all of the observed properties of the system because it does not address the counterrotating structures in IC~0719 or the \hi\ deficiency in IC~0718. IC~0718 is also about a factor of 10 less massive than IC~0719, making it unlikely that IC~0718 could do serious damage to a pre-existing disc in IC~0719.

\section{Discussion}

The primary new result of the \hi\ observations of IC~0719 is that they demonstrate clear and convincing evidence supporting an accretion scenario for the formation of (some of) the $2\sigma$ galaxies.
The warped, irregular disc, which is relaxed into the equatorial plane at small radii, retrograde with respect to the bulk of the stars, and strongly misaligned at large radii, signals that new cold gas was accreted from some external source after the main stellar body was in place.  The faint \hi\ tail pointing towards IC~0718 suggests its identification as the external source.
The angular momentum of the accreted gas was dominated by the geometry of the galaxy interaction and in this case it happened to be misaligned by about 120\arcdeg\ to the rotation of the primary stellar component.
The gas then settled into the equatorial plane in the ``minimal torque" sense, becoming misaligned by 180\arcdeg\ to the primary stellar component.  As the atomic gas condensed into molecular gas, star formation began to form a secondary stellar disc that is counterrotating with respect to the primary stellar disc.

\citet{katkov13} and \citet{pizzella} have conducted extensive optical spectroscopy with analysis of the ionized gas and stellar populations in IC~0719.  They proposed this accretion scenario on the basis of observations such as (1) the primary stellar component is older and more metal rich, and it has a higher velocity dispersion; and (2) the secondary stellar component, the one that rotates like the gas, is younger, with lower metallicity and velocity dispersion.  The fact that the younger stellar component has a lower metallicity clearly suggests the accretion of a large quantity of lower metallicity gas between the star formation episodes.  
The available information about the ionized gas metallicity is also consistent with this interpretation that IC~0719 acquired its gas from IC~0718.  The MPA-JHU catalog \citep[e.g.][]{tremonti}, based on SDSS spectra, quotes $12 + \log(\mathrm{O/H}) = 8.61 \pm 0.06$ for the central regions of IC~0718, and the more detailed studies of \citet{katkov13} and \citet{pizzella} quote values of roughly 8.4 to 8.6 for IC~0719.
But without the new \hi\ data presented here, there was no direct evidence of the warp that smoothly connects the \hi\ bridge of \citet{grossi} through to the dynamically relaxed but retrograde molecular gas on $\sim$ kpc scales.

\subsection{Evolutionary time-scales}

Of the eleven 2$\sigma$ galaxies that have been identified in the \atlas\ survey, only five are currently known to contain large amounts of cold gas (Table \ref{tab:summary}).  Of these five, IC~0719 has the most obvious signs of a misaligned gas disc that is still in the process of settling towards the equatorial plane of the galaxy.  Thus, it is relatively rare to ``catch" a 2$\sigma$ galaxy in this phase of its evolution.

Classically, the torques exerted on gas outside of the equatorial plane should cause the disc to precess and sink towards the equatorial plane in a few dynamical time-scales.
Recent simulations by \citet{freeke_misalign} confirm this idea, with the caveat that one should begin measuring a few dynamical time-scales from the time the accretion of cold gas slows to a trickle, rather than the time that the accretion begins.  During the period of rapid gas accretion, the orbital dynamics of the inflowing gas dominate the angular momentum evolution, and it is only after the accretion shuts down that the torque-driven precession can dominate the evolution and develop a warp.
\citep[][and others, have additionally noted that gas can persist in polar orbits for many dynamical times because the torques are very small.]{bryant2019}

For comparison, in section \ref{sec:hikin} we inferred that the warp in the \hi\ disc of IC~0719 occurs somewhere in the range 15\arcsec\ $\lesssim r \lesssim$ 1\arcmin.   The dynamical time at the outer edge of the CO disc ($r=15$\arcsec) is roughly 70 Myr, at 1\arcmin\ it is 280 Myr, and at the outer edge of the \hi\ disc it is roughly 700 Myr.  The simulations of \citet{freeke_misalign} thus suggest that the gas accretion onto IC~0719 shut down on the order of 0.5 to 2 Gyr ago (6 dynamical times), but not as long as $\sim$ 4 Gyr ago, or a larger portion of the \hi\ disc would be aligned by now.
Of course, this analysis assumes that the dark matter halo of IC~0719 is either spherical or well aligned with the stellar body of the galaxy (so that the outer parts of the \hi\ disc are not in an equatorial plane in their current orientation).
These time-scales are consistent with the finding of \citet{pizzella} that the secondary stellar component has a luminosity-weighted mean age of 1.5 Gyr.

Curiously, there is also evidence from \citet{katkov13} and \citet{pizzella} that there was a previous episode of gas accretion and star formation in IC~0719, beginning on the order of 4 Gyr ago, and also contributing to the growth of the counterrotating (secondary) stellar disc though not dominating its mass.  The timing of that star formation episode is in tension with the timing we infer from the \hi\ kinematics, so it is not clear whether it is connected to the events we now find displayed in the \hi\ properties of the system.

\subsection{Gas and stellar masses}

Based on its $K_s$ luminosity, the stellar mass of IC~0719 is approximately $2\times 10^{10}$ \solmass; we find $10^{9}$ \solmass\ of \hi\ and $2\times 10^8$ \solmass\ of molecular gas (Table \ref{tab:summary}).  Thus, the current masses of \hi\ and \htoo\ are not large enough to form stellar discs that are an appreciable fraction of the current stellar mass, and we infer that the total mass of accreted gas was much higher than the present-day remnants.  Specifically, \citet{pizzella} find that the secondary stellar component in IC~0719 contributes $\sim$ 35\% of the galaxy's light; if it then contains 20\% of the galaxy's stellar mass (because of being younger), its mass is $4\times 10^9$ \solmass.  That is still a factor of two larger than the expected \hi\ content of IC~0718 (Section \ref{sec:0718}).  These considerations do not necessarily rule out IC~0718 as the source of the gas that built the secondary stellar disc, however, because of the unknown contribution of molecular gas, the significant scatter in the cold gas scaling relations \citep{catinella2018}, and because the expected size of the gas reservoir would be galaxies' gas contents several Gyr ago rather than their gas contents today.

Similar comments about the necessity for large accreted masses also apply to NGC~4191, the other \atlas\ 2$\sigma$ galaxy with a relatively large cold gas mass \citep{young4191}.  
For comparison to IC~0719, NGC~4191 has a very large asymmetric \hi\ disc, 110 kpc in diameter, with an internal twist or warp, and again it is clear that the secondary, younger stellar disc formed out of the counterrotating cold gas disc.  Like IC~0719, NGC~4191 also has a gas-rich neighbor nearby in projection (NGC~4180, separated by 600 \kms\ and 160 kpc), but unlike IC~0719, NGC~4191 has no sign of any \hi\ bridge or tail to its neighbor.  \citet{young4191} argued that the gas in NGC~4191 was accreted from an external source based on \hi-stellar misalignments, and \citet{coccato15} used the stellar data to argue for an accretion model through filamentary cold flows.  A filamentary cold flow model has the advantage that it could offer essentially limitless quantities of cold gas, easily explaining the large gas content of NGC~4191.  In short, IC~0719 is similar to NGC~4191 in broad respects but the \hi\ in IC~0719 has much more dramatically disturbed kinematics, perhaps because of being in an earlier dynamical stage or having a different source.

\section{Formation scenarios for the \atlas\ 2$\sigma$ galaxies}

As discussed by \citet{katkov16}, one useful perspective on the formation of the discy stellar counterrotator 2$\sigma$ galaxies 
is that their multiple spin structures required multiple episodes of assembly, and the primary distinction between the various scenarios is how much of the new misaligned or counterrotating material was already in the form of stars when it joined its current host.
In the merger scenario, most of the new material was already in the form of stellar discs; this scenario is rather restrictive in the sense that it requires a special geometry to avoid disrupting the discs during the merger, but it allows a wide variety of ages and velocity dispersions in the final structures.
\citet{bois} show that appropriately aligned mergers with differing mass ratios can make remnants that look like slow rotators with kinematically decoupled cores as well as remnants that look like flattened double disc stellar counterrotators.

The gas accretion scenario accommodates a wider variety of initial geometries, but it makes more specific predictions that one of the discs must be younger and thinner than the other, and that the cold gas must be associated with the thin disc.
This scenario appears to be prevalent in cosmological simulations.
\citet{taylor2018} studied a sample of 82 well-resolved simulated early-type galaxies and found 5 that could be the predecessors of our observed stellar counterrotators, and all of them formed via the accretion mechanism.  \citet{starkenburg2019} also identified the accretion mechanism as the source of long-lived counterrotating gas discs in the galaxies of the Illustris simulation.

IC~0719 is clearly an example of the gas accretion case, but the situation is less clear for the other galaxies observed here.  If we assume that the \hi\ is associated with the thin blue disc and the dust lane in NGC~7710, then NGC~7710 is probably another gas accretion case.  It may be a more advanced version of the scenario in IC~0719, where the kinematic signatures of the accretion are more dissipated, warps have disappeared, and the modest lopsidedness of the \hi\ disc is the only interaction signature remaining in the gas.
But the interpretation of the stellar velocity field in NGC~7710 is not straightforward; if the stellar velocities at large distances above the midplane indicate the rotation of the thick stellar spheroid, then \hi\ rotates like the thicker stellar component, and it's not clear whether that is also the same sense as the blue dusty disc.  The \hi\ certainly does not rotate like the stellar component visible at radii $<$ 4\arcsec\ (Figure \ref{7710vels}).  This latter feature might be a sign of yet a third stellar component in NGC~7710, which complicates its interpretation.
In PGC~056772 the angular resolution in the cold gas data is not good enough to associate the gas with a specific stellar disc, though an extraplanar tail would again suggest the accretion model.  And in both of these cases, detailed decompositions of the stellar populations (like those done for IC~0719 by Katkov et al and Pizzella et al) would be helpful for understanding the relative ages and structural properties of the stellar populations.

Within the context of the accretion scenario, a further distinction might be made between accretion from another galaxy (as appears to be the case for IC~0719) and accretion from a misaligned filament of the cosmic web, as advocated by \citet{algorry}.
If the suggested interaction partner (e.g.\ IC~0718) cannot be identified, it may be difficult to distinguish between these two types of accretion scenarios on the basis of kinematics alone, but measurements of the gas and stellar metallicities should be useful.

Table \ref{tab:summary} summarizes the cold gas contents and proposed formation scenarios for the 2$\sigma$ galaxies in the \atlas\ sample.  We note that in the current data, only one of the eleven 2$\sigma$ galaxies shows dramatic cold gas signatures of interaction or accretion within the last few Gyr.
However, amongst the 2$\sigma$ galaxies there are actually only four that are known to contain any \hi\ at all.  The time-scales discussed above might thus be consistent with the presence of one strongly warped \hi\ disc.  Six of the eleven are currently not detected in either \hi\ or CO emission, and we speculate that they experienced their merger/acquisition processes many Gyr ago and their cold gas has since dissipated.

Of the eleven \atlas\ 2$\sigma$ galaxies, then, five are known to have cold gas; these five, plus an additional one without cold gas detections, have been studied in enough detail to suggest formation mechanisms.  Four of them appear to have formed through a process of clearing out their original cold gas and then acquiring new, misaligned gas that settled into retrograde discs.  The other two appear to have formed through the aligned merger mechanism.  These numbers are, of course, still very small.  More work should be done to develop this specific point of comparison between real galaxies and simulations, as it can be a test of the accuracy of the simulations' treatment of baryonic physics.

Beyond the cases that are presently observed to have two counterrotating stellar discs, the \atlas\ sample contains a number of others with counterrotating molecular discs.  Notable examples in \citet{davis_pv} include NGC~3032, NGC~3626, and NGC~4694.  Still others with counterrotating ionized gas discs are noted in \citet{davis11}, and they may have molecular gas below the detection limits of our CO survey.  The counterrotating molecular disc in NGC~3032 is associated with a small, young counterrotating stellar core which can only be seen with sub-arcsecond stellar kinematic data \citep{mcdermid_conf05,mcdermid_oasis}, so it is clearly a less extreme version of the 2$\sigma$ galaxies, and the others with retrograde atomic and/or molecular discs might be as well.
The \hi\ survey in \citetalias{serra14} also contains many dramatic examples of disturbed \hi\ kinematics in early-type galaxies, and some of these are likely to be in the early stages of misaligned accretion, so they might develop small or faint counterrotating stellar discs in the future \citep{yildiz2017,yildiz2020}.

Thus, the overall importance of the accretion-driven multi-spin phenomenon is likely to be much higher than the $\sim$ 4\% incidence of the 2$\sigma$ galaxies.  This is especially true if one includes the additional 6\% of early-types that are prograde versions of the same phenomenon,  those with counterrotating gas, and possibly also some of the 7\% that have kinematically decoupled cores.

These accretion processes should be expected in spirals as well.
NGC~5719 is a prominent and unusual example of a spiral galaxy showing the double-disc stellar counterrotation phenomenon; it has a
larger \hi\ content than IC~0719, with more dramatic tails \citep{vergani}, and it appears to be in an earlier stage of the process based on the stellar age measurements of \citet{coccato5719}.
The inevitable dissipational nature of \hi\ has led authors such as Coccato et al to suggest that the accretion-driven multi-spin phenomenon can only work in spirals if the mass of accreted gas overwhelms the pre-existing gas.  Small quantities of accreted gas
should simply be entrained into the pre-existing disc on short time-scales.  
Thus one might expect the accretion-driven multi-spin phenomenon to be somewhat easier or more common in early-type galaxies, where the pre-existing cold gas is less abundant.  Along those lines,
more general studies of gas-stellar kinematic misalignments in simulations have confirmed that significant gas loss is an important prior step to developing misalignment \citep{duckworth2019,starkenburg2019}.  
Thus, the early-type galaxies are especially valuable for studying present-day gas accretion rates and their effects on their host galaxies.

\section{Summary}

We present new VLA \hi\ observations of five of the \atlas\ 2$\sigma$ galaxies, plus archival VLA data on another.  These data now complete the cold gas observations of the known 2$\sigma$ galaxies in \atlas such that all of them have been searched for CO emission and all have interferometric \hi\ data.  The detection rate of \hi\ in the 2$\sigma$ galaxies is 4/11, and the detection rate of cold gas (atomic or molecular) is 5/11, similar to that of the \atlas\ sample as a whole \citepalias{serra12}.

We find a small amount of \hi\ in PGC~056772, and the emission is not well resolved so we have little information on its dynamical status.
In NGC~7710 we find an \hi\ disc that is largely relaxed though somewhat lopsided, with 30\% more gas and a larger extent on one side than on the other.  There is no sign of extraplanar gas in NGC~7710 at the current resolution.

IC~0719 shows a strongly warped \hi\ distribution.  The inner part of the \hi\ disc neatly matches the CO distribution and kinematics, which also matches the rotation of the secondary stellar disc but counterrotates with respect to the bulk of the stars.  The outer part of the \hi\ forms an irregular, clumpy disc whose rotation axis is misaligned by about 60\arcdeg\ from the symmetry axis of the galaxy (and whose kinematic position angle is 120\arcdeg\ from the corresponding position angle of the primary stellar component).  A faint \hi\ extension to the south of IC~0719 is probably the bright and close inner portion of a tail connecting IC~0719 to IC~0718, as proposed by \citet{grossi} based on Arecibo data.  The full tail is evidently too faint to be detected in the present VLA data.
IC~0718 shows asymmetric \hi\ and stellar distributions, and a stellar tidal tail, and it is clearly implicated as the source of the \hi\ in IC~0719.
These data therefore confirm a gas accretion model (rather than a binary merger model) for the origin of the counterrotating stellar disc in IC~0719.   The gas fell in along orbits closer to retrograde than to prograde, and as it settled into the symmetry plane it formed the secondary stellar disc.
Estimates of the radius of the \hi\ warp constrain the time-scale of the accretion event such that the heavy period of gas accretion slowed and/or ceased around 0.5 to 2 Gyr ago, in agreement with the stellar ages.

Six of the eleven 2$\sigma$ galaxies in \atlas\ have now been studied in enough detail to make probable identifications of their formation mechanisms.  Two appear to have been formed through the aligned merger process and four through retrograde gas accretion.
Clarifying the physical processes that create the 2$\sigma$ galaxies and their cousins is important because large-scale cosmological simulations are now beginning to reproduce them, so that quantitative comparisons can be used to verify the simulations.  Detailed examination of the \atlas\ sample also suggests that the misalignment/counterrotation phenomenon is very common, with the 2$\sigma$ galaxies being just the most extreme examples of phenomena that are visible in the stars of nearly 20\% of local early-type galaxies.  Misaligned or counterrotating gas is even more common.  These phenomena are not well preserved in spirals, so the early-type galaxies give useful perspectives for comparisons with simulations.

\section*{Acknowledgments}

Thanks to Tim Davis for discussions about the metallicities of IC~0719 and IC~0718.

The National Radio Astronomy Observatory is a facility of the National Science Foundation operated under cooperative agreement by Associated Universities, Inc.

The Legacy Surveys consist of three individual and complementary projects: the Dark Energy Camera Legacy Survey (DECaLS; NOAO Proposal ID \# 2014B-0404; PIs: David Schlegel and Arjun Dey), the Beijing-Arizona Sky Survey (BASS; NOAO Proposal ID \# 2015A-0801; PIs: Zhou Xu and Xiaohui Fan), and the Mayall z-band Legacy Survey (MzLS; NOAO Proposal ID \# 2016A-0453; PI: Arjun Dey). DECaLS, BASS and MzLS together include data obtained, respectively, at the Blanco telescope, Cerro Tololo Inter-American Observatory, National Optical Astronomy Observatory (NOAO); the Bok telescope, Steward Observatory, University of Arizona; and the Mayall telescope, Kitt Peak National Observatory, NOAO. The Legacy Surveys project is honored to be permitted to conduct astronomical research on Iolkam Du'ag (Kitt Peak), a mountain with particular significance to the Tohono O'odham Nation.

NOAO is operated by the Association of Universities for Research in Astronomy (AURA) under a cooperative agreement with the National Science Foundation.

This project used data obtained with the Dark Energy Camera (DECam), which was constructed by the Dark Energy Survey (DES) collaboration. Funding for the DES Projects has been provided by the U.S. Department of Energy, the U.S. National Science Foundation, the Ministry of Science and Education of Spain, the Science and Technology Facilities Council of the United Kingdom, the Higher Education Funding Council for England, the National Center for Supercomputing Applications at the University of Illinois at Urbana-Champaign, the Kavli Institute of Cosmological Physics at the University of Chicago, Center for Cosmology and Astro-Particle Physics at the Ohio State University, the Mitchell Institute for Fundamental Physics and Astronomy at Texas A\&M University, Financiadora de Estudos e Projetos, Fundacao Carlos Chagas Filho de Amparo, Financiadora de Estudos e Projetos, Fundacao Carlos Chagas Filho de Amparo a Pesquisa do Estado do Rio de Janeiro, Conselho Nacional de Desenvolvimento Cientifico e Tecnologico and the Ministerio da Ciencia, Tecnologia e Inovacao, the Deutsche Forschungsgemeinschaft and the Collaborating Institutions in the Dark Energy Survey. The Collaborating Institutions are Argonne National Laboratory, the University of California at Santa Cruz, the University of Cambridge, Centro de Investigaciones Energeticas, Medioambientales y Tecnologicas-Madrid, the University of Chicago, University College London, the DES-Brazil Consortium, the University of Edinburgh, the Eidgenossische Technische Hochschule (ETH) Zurich, Fermi National Accelerator Laboratory, the University of Illinois at Urbana-Champaign, the Institut de Ciencies de l'Espai (IEEC/CSIC), the Institut de Fisica d'Altes Energies, Lawrence Berkeley National Laboratory, the Ludwig-Maximilians Universitat Munchen and the associated Excellence Cluster Universe, the University of Michigan, the National Optical Astronomy Observatory, the University of Nottingham, the Ohio State University, the University of Pennsylvania, the University of Portsmouth, SLAC National Accelerator Laboratory, Stanford University, the University of Sussex, and Texas A\&M University.

The Legacy Surveys imaging of the DESI footprint is supported by the Director, Office of Science, Office of High Energy Physics of the U.S. Department of Energy under Contract No. DE-AC02-05CH1123, by the National Energy Research Scientific Computing Center, a DOE Office of Science User Facility under the same contract; and by the U.S. National Science Foundation, Division of Astronomical Sciences under Contract No. AST-0950945 to NOAO.


\bibliographystyle{mnras}
\bibliography{ic0719}


\appendix

\section{Collateral detections}

\begin{figure}
\includegraphics[width=\columnwidth, clip]{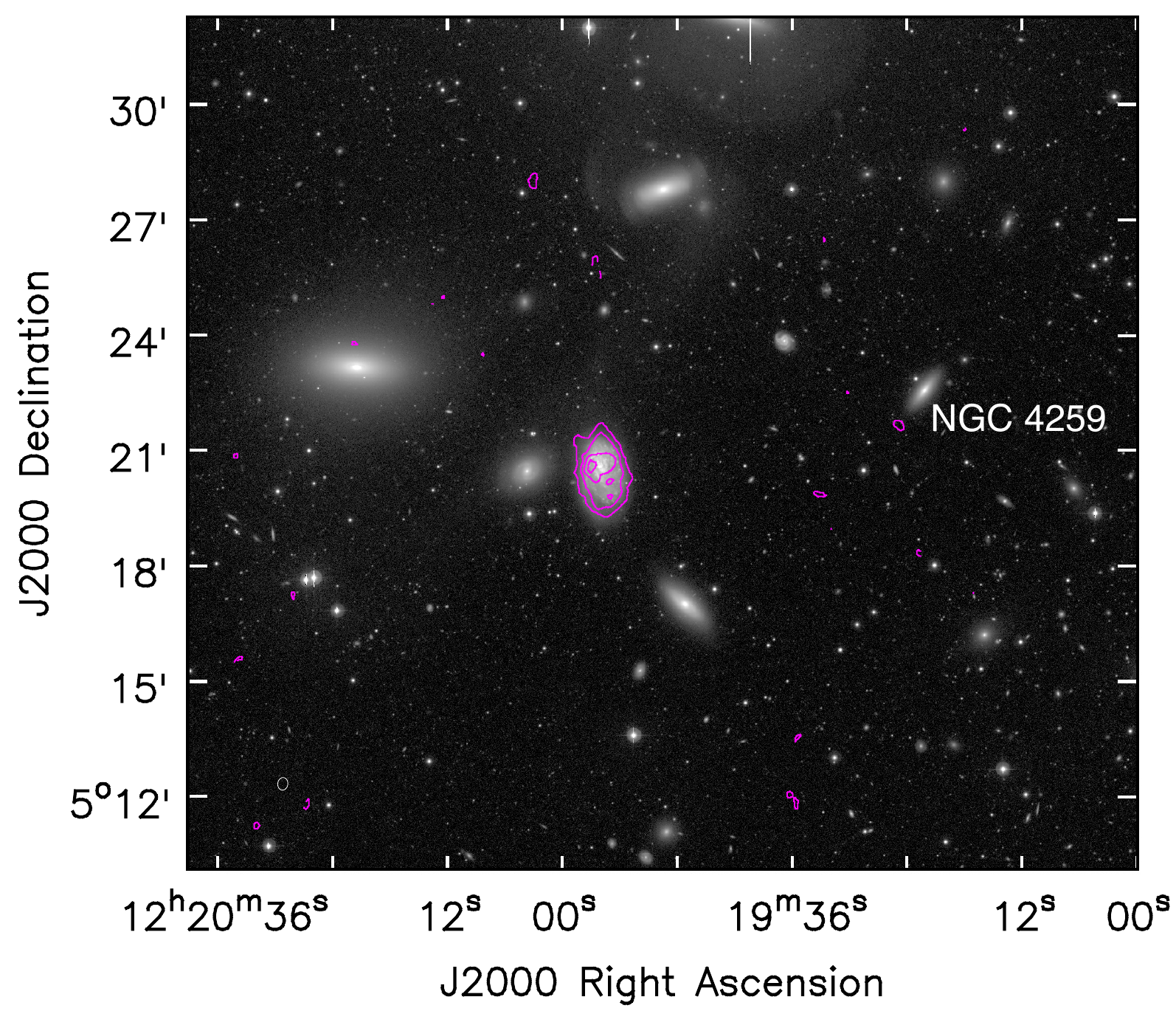}
\caption{\hi\ column density in the field of NGC~4273.  The grayscale is the MATLAS $g$ image \citep{duc15}.  Contours are $(-1, 1, 3, 5, 7)\times 5.4\times10^{20}$\persqcm, which is the nominal column density limit in these data.  The resolution of the \hi\ data is shown as a small white ellipse in the lower left corner.}
\label{4259olay}
\end{figure}

We find no other significant \hi\ emission in the field of NGC~7710 or NGC~0448.  In the field of NGC~4259 there is bright \hi\ emission in the spiral NGC~4273 (Figure \ref{4259olay}); we find 
16.8 $\pm$ 0.8 \jykms\ from NGC~4273, which is rather asymmetric both in \hi\ and in optical emission, but kinematically mostly regular.  Several other galaxies are also within the VLA field of view, but no other \hi\ emission is detected and there are no optical indications that NGC~4259 has suffered a recent interaction.

\begin{figure}
\includegraphics[width=\columnwidth]{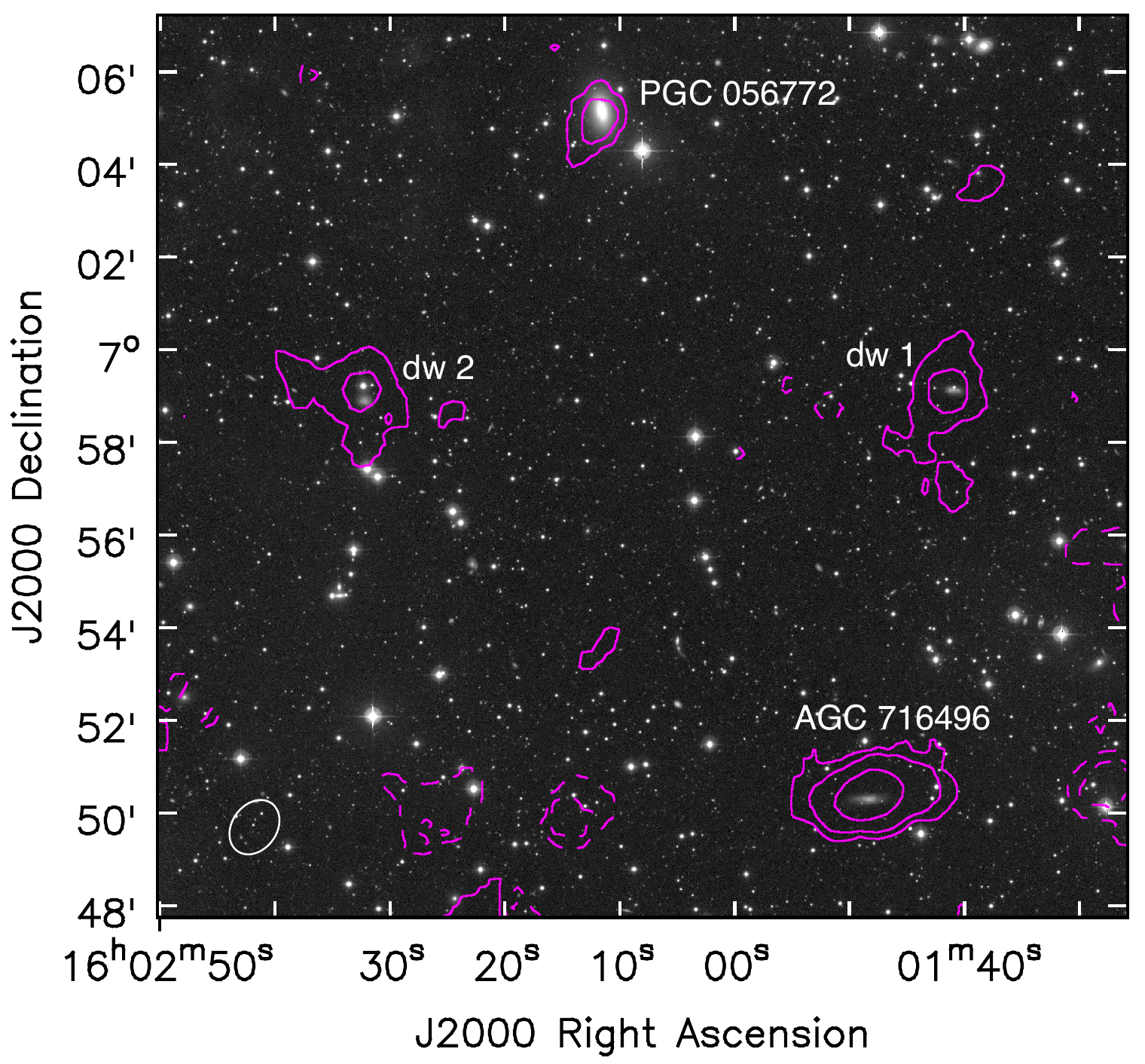}
\caption{\hi\ in the field of PGC~056772.  The greyscale is the MATLAS $g$ image; contours are $(-3, -1, 1, 3, 10)\times 1.36\times 10^{19}$\persqcm, which is the nominal column density limit towards the VLA pointing centre (PGC~056772).  The resolution of the \hi\ data is shown as a white ellipse in the lower left corner.}
\label{pgc056olay}
\end{figure}

\begin{figure}
\includegraphics[width=\columnwidth]{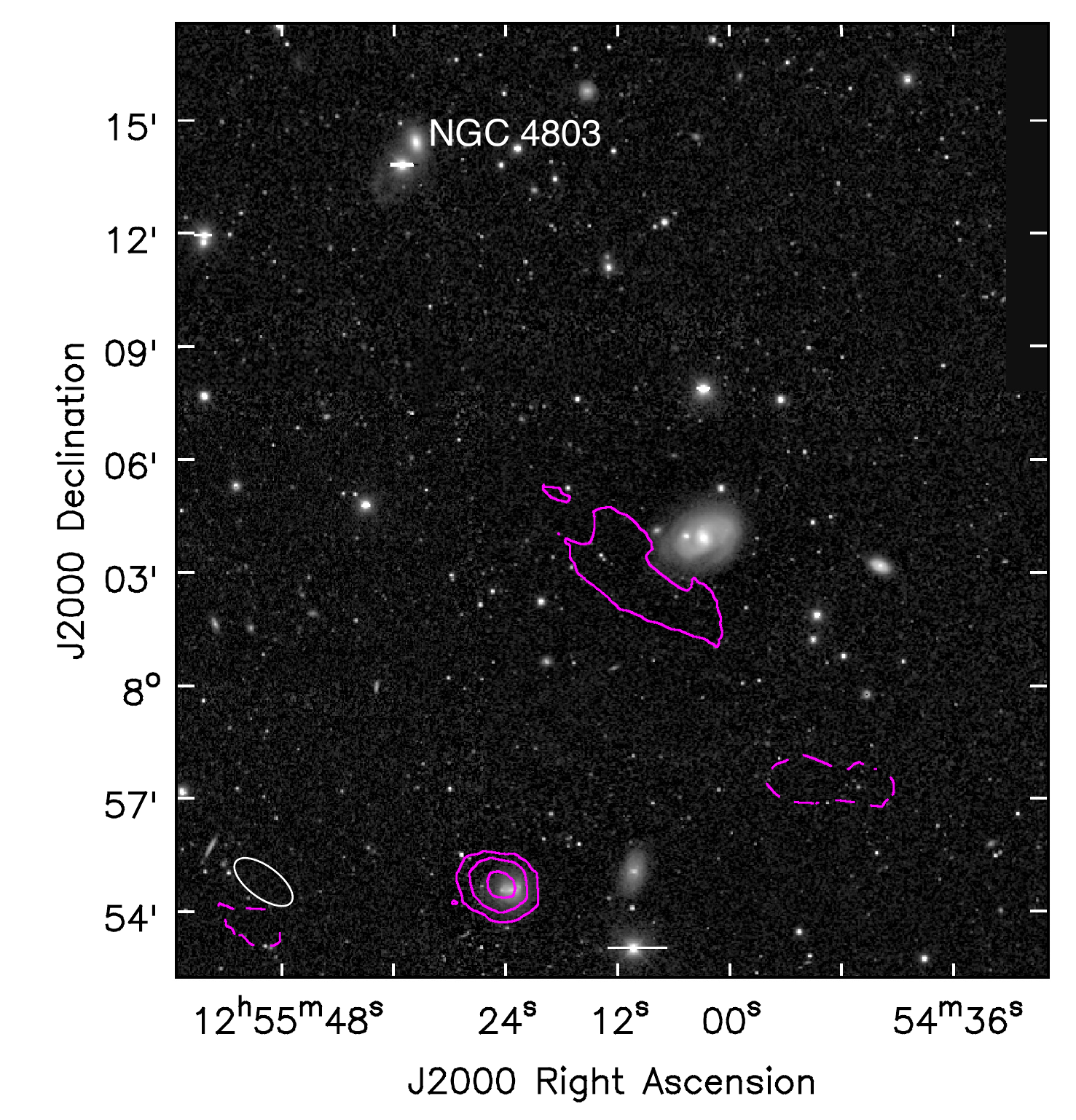}
\caption{\hi\ in the field of NGC~4803.  The greyscale is the DESI Legacy Survey $g$ image; contours are $(-2, 2, 5, 10)\times 1.4\times 10^{19}$\persqcm, which is the nominal column density limit towards the VLA pointing centre (NGC~4803).  An extended contour near the spiral NGC~4795, near the centre of this figure, may represent marginally significant \hi\ emission but as it is near the half-power point of the primary beam, the quality of the imaging is not good.  The resolution of the \hi\ data is shown as a white ellipse in the lower left corner.}
\label{4803olay}
\end{figure}

Our \hi\ observations of PGC~056772 also show emission from three nearby dwarf galaxies (Figure \ref{pgc056olay}).  
The source listed in the ALFALFA catalog as AGC~716496 \citep{alfalfa} has a total flux of 1.20 $\pm$ 0.1 \jykms\ and is found in velocities from 1700 to 1834 \kms.  It is 16.2\arcmin\ from PGC~056772, near the half-power point of the primary beam.  The source we have called ``dwarf 1" has only 0.32 $\pm$ 0.06 \jykms\ in \hi, so it is too faint to be listed in the ALFALFA catalog.  It is found in velocities 1480 -- 1566 \kms.
The source we are calling ``dwarf 2" has 0.19 $\pm$ 0.04 \jykms\ and is also too faint for the ALFALFA catalog, and is found in velocities 2693 -- 2791 \kms.
Finally, in the field of NGC~4803 
we find 1.3 $\pm$ 0.2 \jykms\ of \hi\ emission from UGC~08045, which is 19.6\arcmin\ south of NGC~4803 (Figure \ref{4803olay}).

\bsp	
\label{lastpage}
\end{document}